\documentclass[twocolumn]{aastex631}

\usepackage{color}
\usepackage[utf8]{inputenc}
\usepackage[T1]{fontenc}
\usepackage{amsmath}

\shortauthors{Zhu et al. 2026a}

\begin{document}
	
\title{Common Excitation Patterns of Star Formation, Active Galactic Nuclei, and Shocks in Seyfert Galaxies}

\author[0000-0002-1333-147X]{Peixin Zhu}
\affiliation{Center for Astrophysics $|$ Harvard \& Smithsonian, 60 Garden Street, Cambridge, MA 02138, USA}

\author[0000-0001-8152-3943]{Lisa J. Kewley}
\affiliation{Center for Astrophysics $|$ Harvard \& Smithsonian, 60 Garden Street, Cambridge, MA 02138, USA}
\affiliation{Research School of Astronomy and Astrophysics, Australian National University, Australia}

\author[0000-0002-3626-5831]{Dominika {\L}.~Kr\'{o}l}
\affiliation{Smithsonian Astrophysical Observatory, Center for Astrophysics $|$  Harvard \& Smithsonian, 60 Garden Street, Cambridge, MA 02138, USA}
\affiliation{Astronomical Observatory of the Jagiellonian University, Orla 171, 30-244 Krak\'{o}w, Poland}

\author[0000-0002-3554-3318]{Giuseppina Fabbiano}
\affiliation{Smithsonian Astrophysical Observatory, Center for Astrophysics $|$  Harvard \& Smithsonian, 60 Garden Street, Cambridge, MA 02138, USA}

\author[0000-0001-6950-1629]{Lars Hernquist}
\affiliation{Center for Astrophysics $|$  Harvard \& Smithsonian, 60 Garden Street, Cambridge, MA 02138, USA}

\author[0000-0002-6620-7421]{Ralph S. Sutherland}
\affiliation{Research School of Astronomy and Astrophysics, Australian National University, Australia}

\author[0000-0001-8112-3464]{Anna Trindade~Falcão}
\affiliation{ NASA-Goddard Space Flight Center, Code 662, Greenbelt, MD 20771, USA}
\affiliation{Smithsonian Astrophysical Observatory, Center for Astrophysics $|$  Harvard \& Smithsonian, 60 Garden Street, Cambridge, MA 02138, USA}

\author[0000-0001-5060-1398]{Martin Elvis}
\affiliation{Smithsonian Astrophysical Observatory, Center for Astrophysics $|$  Harvard \& Smithsonian, 60 Garden Street, Cambridge, MA 02138, USA}

\author[0000-0001-9815-9092]{Riccardo Middei}
\affiliation{INAF Osservatorio Astronomico di Roma, Via Frascati 33, 00078 Monte Porzio Catone, RM, Italy}
\affiliation{Smithsonian Astrophysical Observatory, Center for Astrophysics $|$  Harvard \& Smithsonian, 60 Garden Street, Cambridge, MA 02138, USA}

\email{peixin.zhu@cfa.harvard.edu}

\begin{abstract}

The growth of galaxies and their central supermassive black holes is closely connected, yet the net effect of active galactic nuclei (AGN) feedback on host-galaxy star formation remains uncertain. AGN may enhance, suppress, or have little measurable impact on star formation, but distinguishing among these outcomes requires separating star-formation and AGN photoionization from shock excitation, which is expected in AGN-driven outflows but has been difficult to isolate. Here we apply a recently developed theoretical three-dimensional diagnostic diagram, designed to separate star formation, AGN, and shock excitation, to VLT/MUSE IFU observations of nine nearby ($z<0.026$) Type 2 Seyfert galaxies. We find a common excitation pattern across the sample: star-forming rings or arcs at projected radii of $r\sim0.8-6\,$kpc, AGN-photoionized bicones extending to kpc scales, central fast-shock-dominated regions that often extend perpendicular to the AGN bicone, and pure-shock-dominated regions surrounding the central fast shocks and appearing locally within the star-forming rings. Deep Chandra X-ray morphology independently supports this decomposition. The circumnuclear star-forming rings are consistent with bar-driven resonances, although positive AGN feedback may also contribute. The central fast shocks are broadly consistent with AGN jet-ISM interactions, while AGN wind-ISM interactions may also play an important role in galaxies with low-power jets. These results establish central shocks as a common feature of Seyfert galaxies and demonstrate the importance of accounting for shock excitation in AGN feedback studies.

\end{abstract}

\keywords{galaxies: active --- galaxies: ISM --- galaxies: abundances --- ISM: abundance}

\section{Introduction}

Feedback from active galactic nuclei (AGN) has long been proposed to regulate star formation in their host galaxies \citep[e.g.][]{1988ApJ...325...74S,kauffmann_host_2003,wild_bursty_2007,hopkins_cosmological_2008,fabian_observational_2012,heckman_coevolution_2014}. In the positive mode, AGN feedback enhances star formation by compressing the surrounding gas, which then cools and collapses into stars \citep[e.g.][]{2012MNRAS.427.2998I,2013ApJ...772..112S,2013ApJ...774...66Z,cresci_magnum_2015}. In the negative mode, AGN feedback suppresses star formation by heating or expelling the gas that would otherwise fuel it \citep[e.g.][]{2009ApJ...694..599H,muller-sanchez_two_2018,husemann_close_2019}. In semi-analytic models and numerical simulations, AGN feedback is a key ingredient for reproducing the observed properties of massive galaxies \citep[e.g.][]{2005Natur.433..604D,2005MNRAS.361..776S,2006ApJS..163....1H,2010ApJ...717..708C,yuan_active_2018}. Observations of AGN feedback are therefore crucial in constraining the role of supermassive black holes (SMBHs) in galaxy evolution.

However, despite the growing number of observational reports of AGN feedback, its net impact on star formation remains uncertain. Studies based on integrated spectra have found that star formation rates (SFRs) within the central kiloparsec correlate more strongly with SMBH growth rates than galaxy-wide SFRs \citep{2012ApJ...746..168D}, suggesting a positive link between AGN accretion and central starbursts \citep[e.g.][]{kauffmann_host_2003,wild_bursty_2007}. In contrast, more recent spatially resolved studies of individual galaxies show that AGN feedback can vary across a galaxy, with enhanced star formation in some regions \citep{cresci_magnum_2015,shin_positive_2019,falcao_ngc5005_2025} and suppressed star formation in others \citep{husemann_close_2019,shin_positive_2019}. Extending this analysis to a larger sample of 39 AGN host galaxies, \citet{smirnova-pinchukova_close_2022} find no strong evidence for global positive or negative AGN feedback, but only subtle trends suggesting that AGN feedback efficiency may depend on outflow orientation and the star formation history of the host galaxy. Together, these results indicate that AGN feedback operates locally, underscoring the need for spatially resolved studies to unravel its underlying mechanisms.

While spatially resolved studies of large AGN samples are now possible with recent IFU surveys \citep{sanchez_califa_2012,allen_sami_2015,bundy_overview_2015,dopita_probing_2015-1,thomas_probing_2017}, a major challenge that remains is properly accounting for shock emission. Shocks can be generated by AGN feedback when energetic outflows or jets interact with the surrounding interstellar medium (ISM) \citep[e.g.][]{blustin_nature_2005,zakamska_quasar_2014,king_powerful_2015,laha_ionized_2021,riffel_chemical_2021,juneau_black_2022}. Moreover, shocks can significantly contaminate key diagnostics for SFRs (H$\alpha$) and the AGN accretion rate ([\ion{O}{3}]$\lambda5007$) by contributing $\sim35\%$ of the H$\alpha$ flux and up to $\sim50\%$ of the [\ion{O}{3}] flux in nearby Seyfert galaxies \citep{davies_dissecting_2017,sutherland_effects_2017,dagostino_separating_2019, zhu_theoretical_2025}. Properly identifying and separating the contribution from shocks is therefore crucial for reliably assessing the net impact of AGN feedback. 

Despite their physical importance, shocks are often neglected in AGN feedback studies. A key reason is that shocks can produce AGN-like line ratios on standard two-dimensional (2D) optical diagnostic diagrams \citep{baldwin_classification_1981,veilleux_spectral_1987-1,kewley_host_2006}, preventing a clean separation between shock- and AGN-ionized gas \citep{rich_galaxy_2015-1,kewley_understanding_2019,mortazavi_dynamics_2019}. Empirical methods to isolate shocks from AGN and star-formation emission have been proposed \citep{rich_galaxy-wide_2011,ho_sami_2014,medling_shocked_2015,alatalo_shocked_2016,davies_dissecting_2017,dagostino_separating_2019,johnson_empirical_2023}, but they generally suffer from limited applicability and large uncertainties. 

Fortunately, a recently developed theoretical three-dimensional (3D) diagnostic diagram from \citet{zhu_theoretical_2025} overcomes many of these limitations and enables a more reliable separation of star formation, AGN, and shocks. This diagram combines state-of-the-art, self-consistent theoretical models for \ion{H}{2} regions \citep{kewley_understanding_2019}, AGN narrow-line regions (NLRs) \citep{zhu_new_2023}, and time-dependent shocks and precursor models \citep{sutherland_effects_2017,dopita_effects_2017} into a 3D space defined by the emission line ratios [\ion{N}{2}]$\lambda6584$/H$\alpha$, [\ion{O}{3}]$\lambda5007$/H$\beta$, and the [\ion{O}{3}] velocity dispersion ($\sigma_{[\rm O III]}$). Applying galaxy integral field unit (IFU) data to this 3D diagram yields the fractional contributions from star formation, AGN, and shocks for each spaxel, thereby mapping the spatial distributions of \ion{H}{2} regions, AGN NLRs, and shock-excited gas across the galaxy.

Taking advantage of this new 3D diagnostic diagram, this paper presents a pilot study of the spatial distributions of star formation, AGN, and shocks in a sample of nine nearby Seyfert 2 galaxies (including NGC\,5728, as presented in \citet{zhu_theoretical_2025}), using high-resolution optical IFU data and extended X-ray emission from Chandra observations. 

This paper is organized as follows. Section~\ref{sec:data} describes the observational data, and Section~\ref{sec:model} briefly describes the theoretical 3D diagram. Section~\ref{sec:detection} presents the main excitation-separation results, while Section~\ref{sec:result} compares these results with Chandra X-ray observations. We discuss the implications in Section~\ref{sec:discussion} and summarize our conclusions in Section~\ref{sec:conclusion}. Throughout this paper, we adopt a cosmology of $H_0=70\,\rm km\,s^{-1}\,Mpc^{-1}$, $\Omega_{\Lambda}=0.7$, and $\Omega_{m}=0.3$.

\section{Observational Data}\label{sec:data}

As a pilot study, our sample is designed to be representative of nearby Type 2 Seyfert galaxies and is selected based on the availability of public Very Large Telescope (VLT)/MUSE integral field unit (IFU) data and deep ($>$100\,ks) Chandra X-ray observations.  Among the many public optical IFU datasets, MUSE data are preferred because they offer both high spatial sampling (0.2$\arcsec$\,pixel$^{-1}$, {\color{black}although limited by the actual seeing)} and a relatively large field of view (FoV, $1\arcmin\times1\arcmin$), enabling a precise characterization of ionized-gas excitation and kinematics over the central kiloparsec-scale regions where AGN feedback and circumnuclear star formation coexist. The sample comprises nine nearby Type 2 Seyferts: IC 5063, MRK 573, NGC 424, NGC 1386, NGC 3081, NGC 3393, NGC 5643, NGC 5728, and NGC 7212. These galaxies are also favored because they have either HST narrow-band imaging or JWST IFU observations, enabling future higher-spatial-resolution follow-up studies. These galaxies are classified as obscured, Compton-thick AGNs and most show extended soft ($<$3\,keV) X-ray emission in deep Chandra data \citep[for a review, see][]{fabbiano_interaction_2022}. NGC\,424 currently has only a shallow Chandra exposure (46\,ks), but we include it in our sample because deeper Chandra observations (60\,ks, obsid: 30001, 30227) have already been approved. 

We retrieved the MUSE data cubes from the ESO archive. In addition to the fine spatial sampling and relatively large FoV, MUSE provides a spectral resolution of $R\sim1750$–3750 over 4650–9300\,\AA\ \citep{bacon_muse_2010}. For some galaxies, multiple pointings are available, yielding an effectively larger FoV. Table~\ref{tab:1} summarizes the basic properties of each target (RA, Dec, redshift), and the observation dates, total exposure time, seeing, FoV, and program ID of the corresponding MUSE observations. 

\setlength{\tabcolsep}{1.5pt}
\begin{deluxetable*}{c|c|c|c|c|c|c|c|c|c|c}[hbt]
\centering
\tablewidth{1pc}
\tablecaption{Nearby Seyfert 2 Galaxy Sample and MUSE Observing Log\label{tab:1}}
\tablenum{1}
\tablehead{
  \colhead{Galaxy} &
  \colhead{R.A.(J2000)} &
  \colhead{Decl.(J2000)} &
  \colhead{$z$\tablenotemark{\scriptsize a}} &
  \colhead{scale} &
  \colhead{Morphology} &
  \colhead{Observation Dates} &
  \colhead{$t_{\rm exp}$} &
  \colhead{Seeing} &
  \colhead{FoV} &
  \colhead{Program ID}
  \\
  \colhead{} &
  \colhead{(h:m:s)} &
  \colhead{(d:m:s)} &
  \colhead{} &
  \colhead{(pc/$''$)} &
  \colhead{(RC3)\tablenotemark{\scriptsize b}} &
  \colhead{} &
  \colhead{(second)} &
  \colhead{($''$)} &
  \colhead{($'\times'$)} &
  \colhead{}
}
\startdata
IC 5063 & 20:52:02.21 & -57:04:06.96 & 0.011348 & 232 & 
\texttt{SA$0^+$(s)} & 2014/6/23 & 2240 & 0.78 & 1.5$\times$1.5 & 60.A-9339 \\
MRK 573 & 01:43:57.79 & +02:20:58.92 & 0.017179 & 349 & \texttt{(R)SAB$0^+$(rs)} & 2021/1/24 & 4329 & 1.59 & 1.5$\times$1.5 & 106.21C7;110.23PU \\
NGC 424 & 01:11:27.70 & -38:05:00.96 & 0.011764 & 241 & 
\texttt{(R)SB(r)0/a} & 2015/8/13 & 5498 & 0.92 & 1.5$\times$1.5 & 095.B-0934 \\
NGC 1386 & 03:36:46.3 & -35:59:58.00 & 0.003052 & 63 &\texttt{ SB$0^+$(s)} & 2014/11/13 & 3659 & 0.64 & 1.5$\times$1.5 & 094.B-0321 \\
NGC 3081 & 09:59:29.54 & -22:49:34.720 & 0.008149 & 168 & 
\texttt{(R)SAB(r)0/a} & 2017/4/23 & 3279 & 0.71 & 2.1$\times$2.1 & 099.B-0242 \\
NGC 3393 & 10:48:23.40 & -25:09:43.92 & 0.012509 & 256 & \texttt{(R')SB(rs)a:} & 2017/1/31 & 3294 & 0.79 & 1.5$\times$1.5 & 098.B-0551 \\
NGC 5643 & 14:32:40.70 & -44:10:27.84 & 0.003999 & 83 & 
\texttt{SAB(rs)c} & 2015/5/12 & 3288 & 0.56 & 1.5$\times$1.5 & 095.B-0532 \\
NGC 5728 & 14:42:23.90 & -17:15:11.16 & 0.009353 & 192 & 
\texttt{SAB(r)a:} & 2016/4/3 & 4747 & 0.69 & 1.6$\times$1.6 & 097.B-0640 \\
NGC 7212 & 22:07:02.08 & +10:14:03.154 & 0.026632 & 535 & interacting\tablenotemark{\scriptsize c} & 2023/11/25 & 2296 & 1.23 & 1.8$\times$1.8 & 112.25VE \\
\enddata
\tablenotetext{\scriptsize a}{Redshift obtained from https://ned.ipac.caltech.edu/. }
\tablenotetext{\scriptsize b}{Morphological classifications are adopted from the Third Reference Catalogue of Bright Galaxies \citep[RC3;][]{1991rc3..book.....D}. Here \texttt{(R)} denotes an outer ring, \texttt{(R')} represents an outer pseudo-ring. \texttt{SA}, \texttt{SAB}, and \texttt{SB} indicate unbarred, weakly barred, and strongly barred systems, respectively. The inner structure is encoded by \texttt{(r)} (inner ring) and \texttt{(rs)} (intermediate ring/spiral morphology). The stage codes describe Hubble type: \texttt{0$^{+}$} denotes a late-type lenticular (S0$^{+}$); \texttt{0/a} marks the transition between S0 and Sa; \texttt{a} indicates an early-type spiral with tightly wound arms; and \texttt{c} indicates a late-type spiral with loosely wound arms. ``:" indicates that the classification is uncertain.}
\tablenotetext{\scriptsize c}{NGC\,7212 lies at the center of a three-galaxy interacting system.}
\end{deluxetable*}
\vspace{-2.em} 

All MUSE data cubes are first processed with the nGIST pipeline \citep{bittner_gist_2019,fraser_ngist_2025,fraser_geckos_2025}\footnote{\url{http://ascl.net/1907.025}; \url{https://ascl.net/2507.015}}, which performs stellar continuum subtraction and single-Gaussian emission-line fitting. The continuum is modeled using a combination of single stellar population (SSP) templates from the MILES library \citep{vazdekis_evolutionary_2010}, and the emission lines are fitted with the penalized pixel-fitting (pPXF) code \citep{cappellari_parametric_2004,cappellari_improving_2017}. Consistent with previous work \citep[e.g.,][]{cresci_magnum_2015,mingozzi_magnum_2019,venturi_magnum_2021}, we find that spaxels in the central regions of some galaxies exhibit kinematically complex line profiles, leaving significant residuals when modeled with a single Gaussian.

 For these spaxels, we refit the continuum-subtracted spectra with pPXF using up to three Gaussian components, selecting the preferred model based on the reduced $\chi^2$. In both the single- and multi-Gaussian fits, emission lines are divided into three groups: (i) low-ionization lines ([N~II], [N~I], [O~I], [S~II]); (ii) high-ionization lines ([O~III], He~I); and (iii) Balmer lines (H$\alpha$, H$\beta$). Within each group, the velocity ($v$) and velocity dispersion ($\sigma$) are tied to share the same kinematic parameters. For each galaxy, our analysis only includes spaxels with signal-to-noise (S/N) $>3$ in all of the [\ion{N}{2}]$\lambda6584$, H$\alpha$, [\ion{O}{3}]$\lambda5007$, and H$\beta$ emission-line fluxes.
 
We also include X-ray data from archival Chandra X-ray observations. These AGNs are included in the sample recently reprocessed uniformly by R. Middei et al. (2026, submitted). They all have highly obscured nuclear sources, making it possible to obtain good images of the extended X-ray emission of their ISM. This X-ray emission is typically due to both photoionization from the AGN and collisional excitation that may be connected to shocks from the interaction of radio jets and high-velocity nuclear winds with the galaxy ISM \citep[see the review by][]{fabbiano_interaction_2022}. 

Co-added Chandra ACIS-S exposures longer than 100 ks are available for eight out of nine AGNs in this study, and will be available for NGC\,424. {\color{black}The Chandra datasets used in this paper are contained in \dataset[DOI: 10.25574/cdc.661]{https://doi.org/10.25574/cdc.661}.} The data were co-added and analyzed spatially as described in R. Middei et al. (2026, submitted), using CIAO tools \citep{ciao_2006}. XSPEC \citep{xspec_1996} was used for the spectral analysis. 

{\color{black} To construct X-ray contours in a given energy band, we first filter the Chandra images to that band and then apply $1/8$ sub-pixel binning to exploit the full $\sim0.3\arcsec$ Chandra mirror resolution. We then derive the contours from adaptively smoothed images generated with the CIAO task \texttt{dmimgadapt}, using parameters \texttt{func=gaussian min=2 max=15 num=30 radscale=log counts=5}. Here \texttt{func=gaussian} specifies Gaussian smoothing; \texttt{min=2}, \texttt{max=15}, \texttt{num=30}, and \texttt{radscale=log} define 30 logarithmically spaced smoothing scales from 2 to 15 image pixels; and \texttt{counts=5} requires each kernel to enclose at least five counts. The resulting images therefore have position-dependent effective spatial resolution: compact, high-surface-brightness structures are smoothed on smaller scales, whereas fainter diffuse emission is smoothed on larger scales, up to the maximum scale of 15 pixels.} We truncate the X-ray contours at the sky background level, which is estimated from background-dominated regions identified in the unbinned X-ray image as areas with counts below 3$\sigma$ of the background fluctuations.

\section{Theoretical 3D diagnostic diagram}\label{sec:model}

We refer to \citet{zhu_theoretical_2025} for a full description of the theoretical 3D diagnostic diagram and provide only a brief overview here. The diagram extends the classical 2D Baldwin-Phillips-Terlevich (BPT) diagram \citep{baldwin_classification_1981,veilleux_spectral_1987-1,kewley_host_2006} by using the emission-line ratios [\ion{N}{2}]$\lambda6584$/H$\alpha$ and [\ion{O}{3}]$\lambda5007$/H$\beta$ as the first two axes, and adding the [\ion{O}{3}]$\lambda5007$ velocity dispersion ($\sigma_{[\rm O III]}$) as the third axis to maximize the separation between \ion{H}{2} regions, AGN, and shocks. Theoretical models for \ion{H}{2} regions, AGN NLRs, fast shocks, and pure shocks are embedded in this 3D space to facilitate excitation-source classification, {\color{black}which is necessary because} the line ratios depend not only on ISM conditions but also on source properties such as the AGN SED shape \citep{zhu_new_2023} and shock velocity \citep{sutherland_effects_2017}. 

The theoretical models used in the 3D diagram are isobaric \ion{H}{2} region photoionization models from \citet{kewley_understanding_2019}, isobaric AGN NLR photoionization models from \citet{zhu_new_2023}, and shock and precursor models from \citet{sutherland_effects_2017,dopita_effects_2017}. These models are all calculated with MAPPINGS version 5.2 \citep{binette_radiative_1985,sutherland_cooling_1993,dopita_new_2013,sutherland_mappings_2018}, using atomic data from the CHIANTI version 10 database \citep{del_zanna_chiantiatomic_2021} for the 30 lightest elements. Abundance scaling relations from observations of nearby \ion{H}{2} regions from \citet{2017MNRAS.466.4403N} are used in these models, which better characterize element abundances in O and B stars than uniform solar-based scaling relations. The effects of dust depletion are included in all models using the factors from \citet{jenkins_unified_2009}. For shock models, dust destruction is also included because dust is known to be largely destroyed by shocks \citep{dopita_spectral_1995,beck-winchatz_gas-phase_1996,reipurth_hubble_2000}.

The stellar ionizing spectra in \ion{H}{2} region photoionization models are generated with Starburst99 \citep{leitherer_starburst99_1999,leitherer_effects_2014}. These stellar ionizing spectra are calculated using the Salpeter initial mass function \citep{salpeter_luminosity_1955} with an upper mass limit of 100$\,M_{\odot}$, the Pauldrach/Hillier stellar atmosphere model \citep{hillier_treatment_1998,pauldrach_radiation-driven_2001}, the Geneva group ``high'' mass-loss evolutionary tracks from \citet{meynet_grids_1994}, and assuming a continuous star-formation history. The stellar ionizing spectra are obtained at a stellar age of 5 Myr. 
Variable parameters in the \ion{H}{2} region models are gas-phase metallicity ($7.3\leq$12+$\log(\rm O/H)\leq9.4$), ionization parameter ($-4.0\leq\log(U)\leq-1.0$), and gas pressure ($5.2\leq\log{(P/k)}\leq7.8$). {\color{black}These parameter ranges are chosen to encompass the values observed in local \ion{H}{2} regions \citep{kewley_understanding_2019}.}

The AGN ionizing spectra in AGN NLR photoionization models are generated from OXAF \citep{thomas_physically_2016}, a simplified version of the AGN radiation model OPTXAGNF \citep{done_intrinsic_2012,jin_combined_2012-1} that describes the continuum emission radiated from an AGN thin accretion disk and a Comptonizing corona surrounding a central rotating black hole. $E_{\rm peak}$ is the most important parameter in the OXAF model, which characterizes the peak energy of the SED, dominated by a \citet{shakura_black_1973} accretion disk thermal emission, and is determined jointly by the black hole mass and Eddington accretion rate. Within its representative range $-2.0\leq\log (E_{\rm peak}/\rm keV)\leq-1.0$, $E_{\rm peak}$ can change the predicted emission-line ratios by $\gtrsim0.5\,$dex \citep{thomas_physically_2016,zhu_new_2023}. Variable parameters in the AGN NLR models are $E_{\rm peak}$ ($-2.0\leq\log (E_{\rm peak}/\rm keV)\leq-1.0$), gas-phase metallicity ($7.3\leq$12+$\log(\rm O/H)\leq9.4$), ionization parameter ($-4.0\leq\log(U)\leq-1.0$), and gas pressure ($6.2\leq\log{(P/k)}\leq7.8$). These parameter ranges are chosen to encompass the values observed in nearby AGNs \citep{zhu_new_2023}. 

The radiative shock and precursor models described in \cite{sutherland_effects_2017,dopita_effects_2017} are, for the first time, calculated using fully self-consistent treatments by applying a multi-zone iterative scheme to solve the time-dependent photoionization, recombination, photoelectric heating, and line cooling processes throughout the preshock and postshock regions. The multi-zone iterative scheme provides more sophisticated and realistic predictions of shock and precursor emission than previous shock models. The shock velocity ($v_s$) is the most crucial parameter in shock models, as it determines the postshock temperature, the postshock radiation field, and the ionized states of the precursor (preshock gas). The variable parameters in the shock and precursor models are shock velocity $v_s$ ($150\leq v_s (\rm km/s)\leq1500$), gas-phase metallicity ($7.3\leq$12+$\log(\rm O/H)\leq9.4$), gas pressure ($6.2\leq\log{(P/k)}\leq10.2$), and the magnetic-to-ram pressure ratio\footnote{$\eta_M=B_0^2/(4\pi\rho_0v_s^2)$, where $B_0$ is the transverse component of the preshock magnetic field and $\rho_0$ is the mass density of the preshock gas.} ($0\leq\eta_M\leq0.1$, corresponding to a magnetic field strength in the range $0\leq B_0\lesssim1.721\times10^3\mu G$). These parameter ranges are chosen to be representative of those measured in nearby AGN-driven shocks \citep{sutherland_effects_2017,zhu_theoretical_2025}.

To characterize AGN-driven shocks, we restrict our attention to models with $v_s\gtrsim150\,$km/s. In this regime, the precursor is fully ionized, has reached photoionization equilibrium, and can contribute significantly to the total shock emission \citep{sutherland_effects_2017}. However, as shocks propagate through an inhomogeneous ISM, the precursor may be incomplete in some regions. We therefore include two classes of shock models in the theoretical 3D diagram to represent different levels of precursor contribution: ``pure shock" models and ``fast shock" models. The pure shock models are dominated by postshock emission, with little to no precursor contribution ($\lesssim 5\%$), whereas the fast-shock models include a substantial precursor component ($\gtrsim 30\%$). When the precursor contribution is non-zero, the total shock spectrum is obtained by summing the shock and precursor components at a fixed ratio of their H$\beta$ luminosities ($L_{\mathrm{H}\beta,\mathrm{sh}} : L_{\mathrm{H}\beta,\mathrm{pre}}$). This ratio may vary from galaxy to galaxy because it depends on the local ISM structure and on the detailed interaction between AGN-driven outflows or jets and the surrounding gas. We therefore decide these ratios empirically, guided by the locations of the observed shock sequences in the theoretical 3D diagnostic diagrams. The adopted ratios for each galaxy are listed in Table~\ref{tab:3}.

To place these models on the 3D diagram, we assign the \ion{H}{2} and AGN model grids a characteristic velocity dispersion based on the empirical mean $\sigma_{[\rm O III]}$ measured in the corresponding \ion{H}{2}-dominated and AGN-dominated spaxels. These spaxels are selected using the \citet{kauffmann_host_2003} and \citet{kewley_theoretical_2001} demarcation lines on the BPT diagram, respectively, together with upper limits of $\sigma_{[\rm O III]}\leq120\,\rm km\,s^{-1}$ for the \ion{H}{2}-dominated spaxels and $\sigma_{[\rm O III]}\leq400\,\rm km\,s^{-1}$ for the AGN-dominated spaxels. The resulting model velocity dispersions are summarized in Table~\ref{tab:3}. For the shock models, we adopt the shock velocity ($v_s$)  as a proxy for $\sigma_{[\rm O III]}$, assuming an upright viewing geometry. While projection effects and shock instabilities may reduce the observed $\sigma_{[\rm O III]}$, we find that this approximation provides a reasonable description for the majority of our sample.

\section{Excitation-Source Decomposition: \ion{H}{2}, AGN, and Shocks}\label{sec:detection}

This section presents the theoretical 3D diagnostic diagram and the 2D excitation-source decomposition maps for star formation, AGN, fast shocks, and pure shocks (if present) in eight galaxies: IC\,5063, MRK\,573, NGC\,424, NGC\,1386, NGC\,3081, NGC\,3393, NGC\,5643, and NGC\,7212. The excitation-source decomposition for NGC\,5728 has already been presented in \citet{zhu_theoretical_2025}, which we will adopt directly in the X-ray comparison. 

We present the theoretical 3D diagram for IC\,5063 as a representative example in the main text in Figure~\ref{fig:5063_3d}, with the corresponding theoretical 3D diagrams for the remaining galaxies are presented in Appendix~\ref{sec:appendixA}. These 3D diagrams are projected at viewing angles chosen to highlight the shock sequences, although these projections make the \ion{H}{2}--AGN mixing sequence less apparent. The interactive versions available in the online journal provide a clearer view of the full 3D distribution of the IFU data and theoretical model grids.
 
Using the theoretical 3D diagnostic diagram, we follow the methodology of \citet{zhu_theoretical_2025} to estimate the fractional contributions from \ion{H}{2} regions, AGN, fast shocks, and pure shocks, along with their uncertainties, for each spaxel. For spaxels that contain more than one Gaussian component, each component is treated independently in the 3D diagram because the components exhibiting different kinematics may arise from distinct physical regions and be excited by different sources. In the 2D maps, we sum the fractional contributions of all Gaussian components belonging to the same excitation source, allowing spaxels with multi-Gaussian components to more clearly reflect the presence of that source.

\begin{figure*}[hbt]
\begin{interactive}{js}{IC5063ppxf_int.zip}
\epsscale{1.13}
\plotone{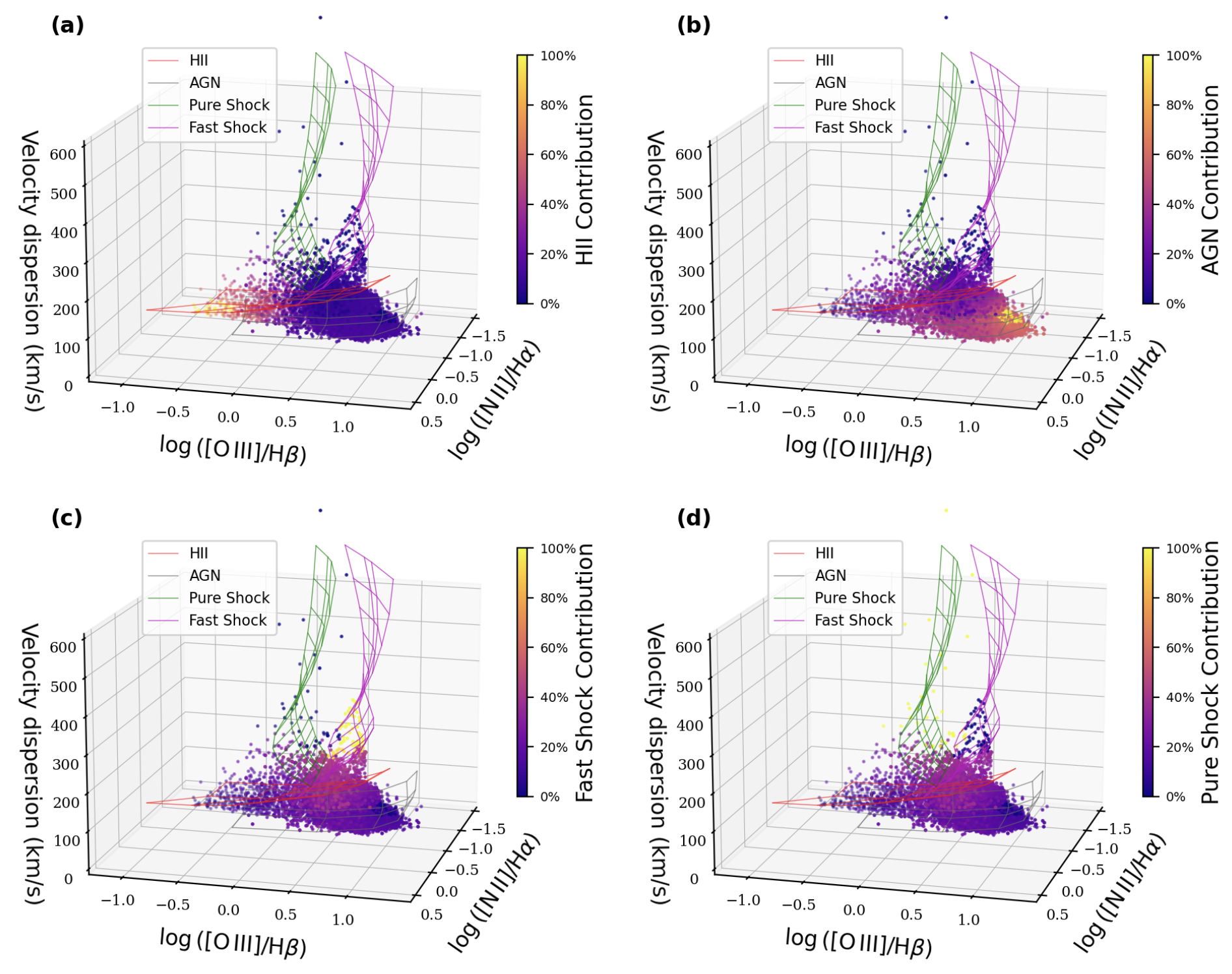}
\end{interactive}
\caption{Distribution of MUSE IFU data for IC\,5063 on the theoretical 3D diagram, with spaxels color-coded by the fractional contributions of \ion{H}{2} regions (top left), AGN (top right), fast shocks (bottom left), and pure shocks (bottom right). The \ion{H}{2} models with $\log({P/k})=6.0$, AGN models with $\log({P/k})=7.4$ and $\log E_{\text{peak}}/(\text{keV})=-1.0$, fast-shock, and pure-shock model grids with $\log({P/k})=10.2$ and $\eta_M=0.0001$ are overplotted in red, grey, magenta, and green, respectively, showing lines of constant metallicity ($12+\log(\rm O/H)=8.12, 8.42, 8.82$ for HII model; $12+\log(\rm O/H)=8.43, 8.70, 8.80, 9.02, 9.26$ for AGN and shock models), constant ionization parameter ($\log(\rm U)=-3.75$ to $-2.25$  in steps of 0.25 dex for the HII model and $\log(\rm U)=-3.8$ to $-2.2$ in steps of 0.4 dex for the AGN model), and constant shock velocity ($V_s=158, 179, 202, 228, 257, 290, 328, 370, 418, 472, 533, 601, 678, 766\,\rm km\,s^{-1}$) for the shock models. The x- and y-axis ranges are tailored to encompass the full range of IFU data and theoretical models, as shown clearly in the interactive version of this figure, available in the online journal. 
\label{fig:5063_3d}}
\end{figure*}

\begin{figure*}[hbt]
\epsscale{1.21}
\plotone{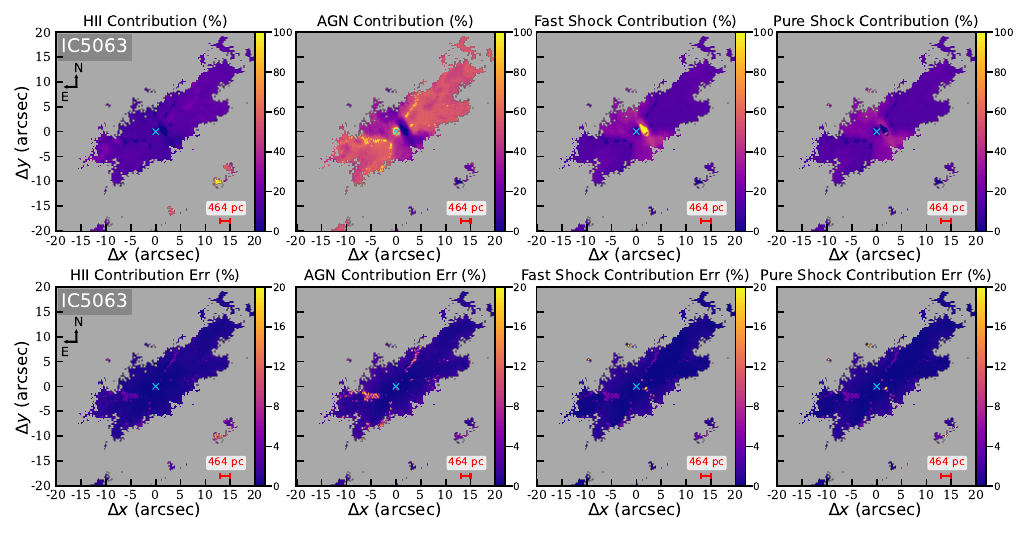}
\caption{2D maps of the fractional contributions (top panel) and their uncertainties (bottom panel) for IC\,5063. The spaxel with the highest [\ion{O}{3}] luminosity, marking the galaxy center, is indicated by a cyan cross.
\label{fig:5063_2d}}
\end{figure*}

\begin{figure*}[hbt]
\epsscale{1.21}
\plotone{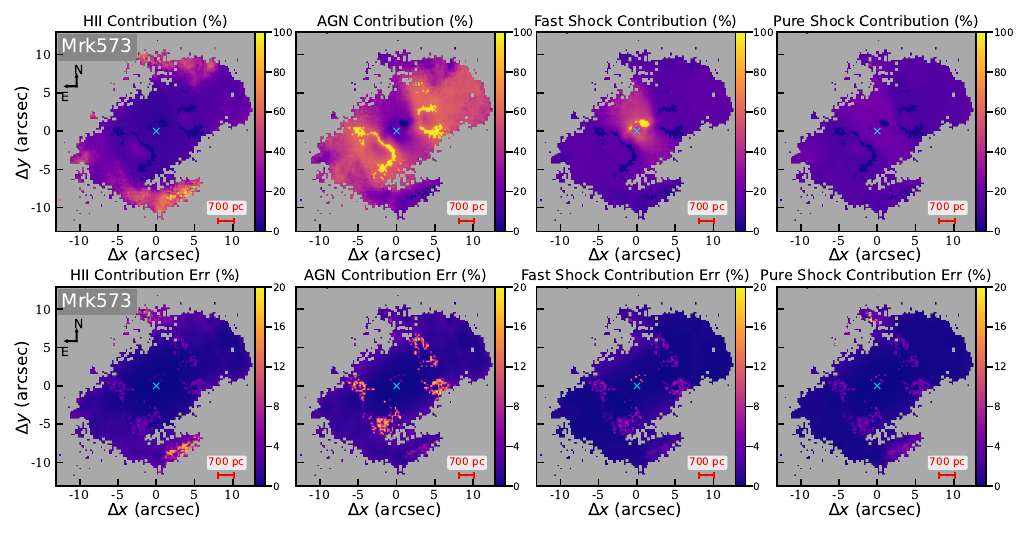}
\caption{2D maps of the fractional contributions (top panel) and their uncertainties (bottom panel) for Mrk\,573. The panel layout and model grid parameters are the same as in Figure~\ref{fig:5063_2d}.
\label{fig:573_2d}}
\end{figure*}

\begin{figure*}[hbt]
\epsscale{1.21}
\plotone{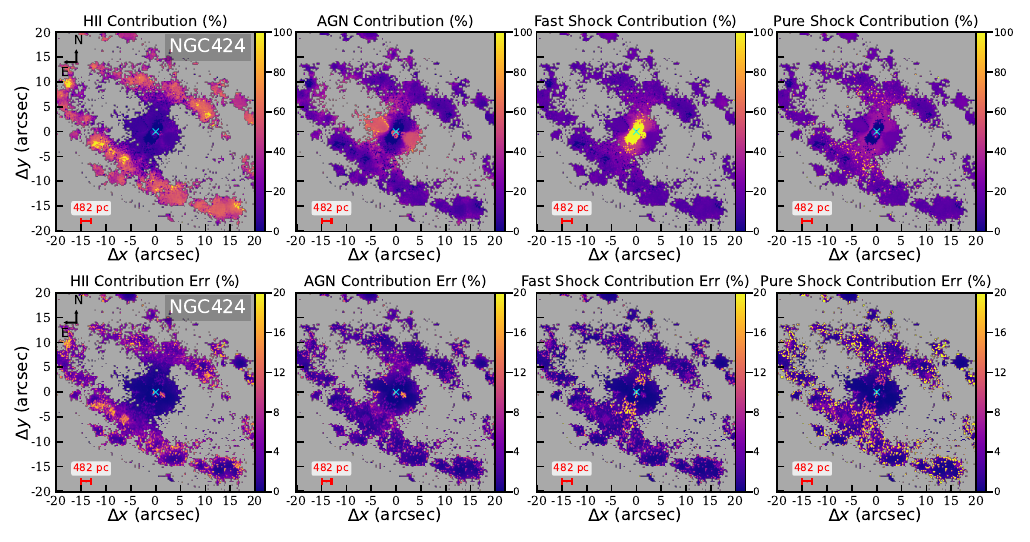}
\caption{2D maps of the fractional contributions (top panel) and their uncertainties (bottom panel) for NGC\,424. The panel layout and model grid parameters are the same as in Figure~\ref{fig:5063_2d}. \label{fig:424_2d}}
\end{figure*}

Figure~\ref{fig:5063_3d} presents the theoretical 3D diagram with IFU data of IC\,5063 color-coded by the estimated fractional contribution. In the 3D diagram, the IC\,5063 spaxels trace a low velocity dispersion \ion{H}{2}$-$AGN mixing sequence, together with two nearly vertical sequences indicating \ion{H}{2}$-$pure shock mixing and AGN$-$fast shock mixing. This distribution indicates that the MUSE IFU data of IC\,5063 contain four excitation sources. The corresponding 2D fractional maps are shown in Figure~\ref{fig:5063_2d}, where \ion{H}{2}-dominated spaxels are sparsely distributed at projected distances of $\sim4\,$kpc from the nucleus. The AGN-dominated emission forms two elongated ionization cones, each extending to $\sim3.5-5.3\,$kpc from the nucleus with a characteristic width of $\sim2.3\,$kpc, oriented toward the northwest and southeast. Fast shocks dominate a region $\sim250\,$pc northwest of the galaxy center and extend up to $\sim1\,$kpc perpendicular to the AGN ionization cones, with enhanced pure-shock contributions surrounding the fast-shock-dominated zone. This distribution indicates that a pure shock could result from a fast shock losing its precursor emission due to ISM inhomogeneity. 

For MRK\,573, the theoretical 3D diagram in Figure~\ref{fig:573_3d} likewise indicates the presence of all four excitation sources. Most spaxels lie along the low velocity dispersion \ion{H}{2}$-$AGN mixing sequence, while a small subset populates regions of the diagram associated with the pure shock and fast shock model grids. The 2D fractional maps in Figure~\ref{fig:573_2d} show that the \ion{H}{2}-dominated emission primarily originates from an incomplete star-forming ring with a projected radius of $\sim3\,$kpc from the nucleus, detected at sufficient signal-to-noise in the northern and southern portions of the ring. \ion{H}{2}-dominated emission toward the northeast is apparent in a separate, longer-exposure MUSE dataset, but we do not use that cube here because its poorer seeing ($\approx$ 2.3\arcsec) would compromise the spatial resolution. The AGN-dominated spaxels delineate two elongated ionization cones extending to the northwest and southeast, with characteristic dimensions of $\sim3\,$kpc in length and $\sim3.5\,$kpc in width. The fast-shock component dominates the central regions ($r\lesssim700\,$pc) and extends up to $\sim1.8\,$kpc, perpendicular to the AGN ionization cones, with enhanced pure-shock contributions surrounding the central fast-shock region.

Similarly, NGC\,424 exhibits contributions from all four excitation sources in the theoretical 3D diagram, with clear loci corresponding to \ion{H}{2}$-$AGN mixing, \ion{H}{2}$-$pure shock mixing, and AGN$-$fast shock mixing sequences in Figure~\ref{fig:424_3d}. The 2D fractional maps in Figure~\ref{fig:424_2d} show that the \ion{H}{2}-dominated spaxels are associated with a star-forming ring with a projected semimajor axis of $\sim4.8\,$kpc. AGN-dominated spaxels delineate two ionization cones oriented toward the northeast and southwest, extending to $\sim1\,$kpc from the nucleus in the MUSE data. Emission at larger radii along the cone directions falls below the MUSE signal-to-noise threshold, but deeper S7 observations detect and confirm AGN-dominated emission on these scales (P. Zhu et al. 2026c, in prep). Fast shock-dominated emission is concentrated near the nucleus, at the base of the AGN ionization cones. Pure-shock contributions are enhanced both around this central fast-shock region and, more sparsely, within the star-forming ring, suggesting that pure shocks could also arise from stellar feedback.

\begin{figure*}[hbt]
\epsscale{1.22}
\plotone{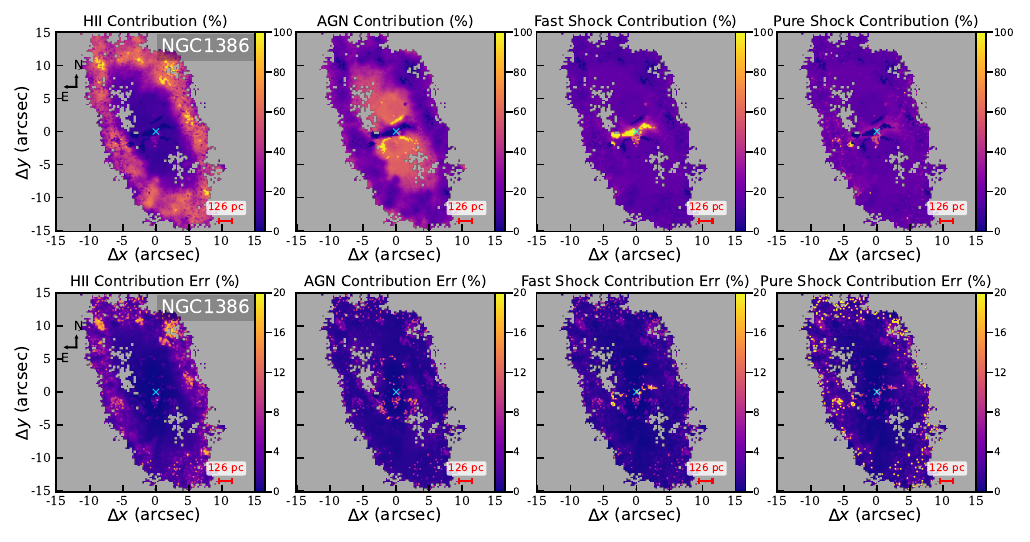}
\caption{2D maps of the fractional contributions (top panel) and their uncertainties (bottom panel) for NGC\,1386. The panel layout and model grid parameters are the same as in Figure~\ref{fig:5063_2d}. 
\label{fig:1386_2d}}
\end{figure*}

\begin{figure*}[hbt]
\epsscale{1.21}
\plotone{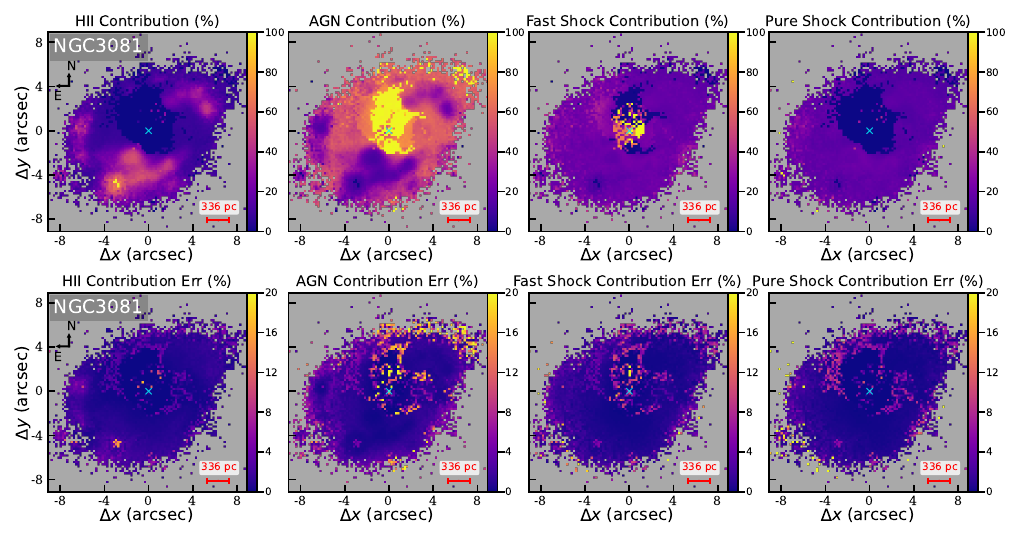}
\caption{2D maps of the fractional contributions (top panel) and their uncertainties (bottom panel) for NGC\,3081. The panel layout and model grid parameters are the same as in Figure~\ref{fig:5063_2d}.
\label{fig:3081_2d}}
\end{figure*}

\begin{figure*}[htb]
\epsscale{1.2}
\plotone{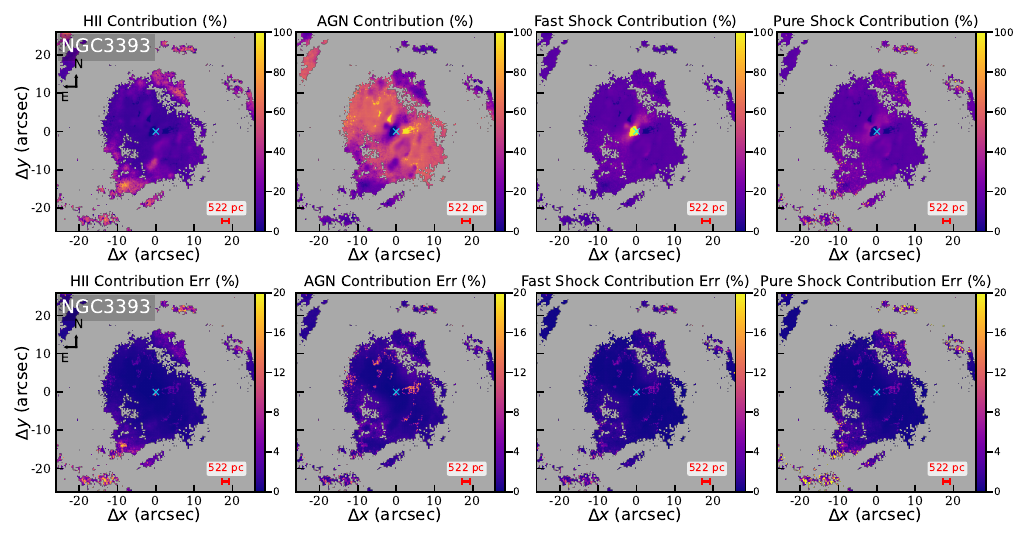}
\caption{2D maps of the fractional contributions (top panel) and their uncertainties (bottom panel) for NGC\,3393. The panel layout and model grid parameters are the same as in Figure~\ref{fig:5063_2d}.
\label{fig:3393_2d}}
\end{figure*}

\begin{figure*}[hbt]
\epsscale{1.22}
\plotone{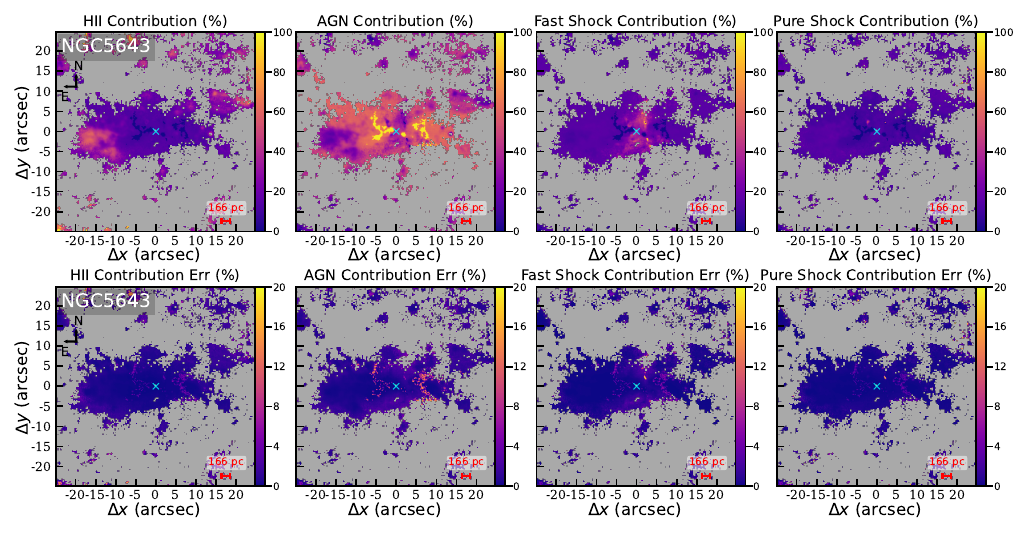}
\caption{2D maps of the fractional contributions (top panel) and their uncertainties (bottom panel) for NGC\,5643. The panel layout and model grid parameters are the same as in Figure~\ref{fig:5063_2d}.
\label{fig:5643_2d}}
\end{figure*}

\begin{figure*}[hbt]
\epsscale{1.1}
\plotone{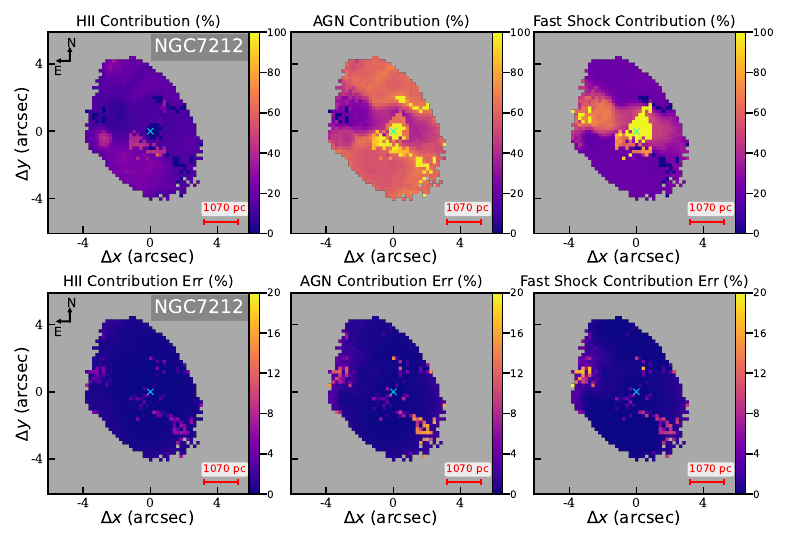}
\caption{2D maps of the fractional contributions (top panel) and their uncertainties (bottom panel) for NGC\,7212. The panel layout and model grid parameters are the same as in Figure~\ref{fig:5063_2d}.
\label{fig:7212_2d}}
\end{figure*}

The theoretical 3D diagnostic diagrams in Figure~\ref{fig:1386_3d} indicate that the MUSE data for NGC\,1386 likewise contain all four excitation sources. As shown by the 2D fractional maps in Figure~\ref{fig:1386_2d}, the \ion{H}{2}-dominated spaxels trace a compact nuclear star-forming ring with a projected semimajor axis of $\sim800\,$pc. AGN-dominated spaxels delineate two ionization cones extending along the north-south direction (each $\sim700\,$pc in length and $\sim500\,$pc in width) and are enclosed by the star-forming ring. The shock distribution resembles that in NGC\,424: fast shocks are concentrated in the central regions, while pure shocks form an encircling structure around the fast-shock-dominated zone and also appear more sparsely within the star-forming ring.

NGC\,3081 also exhibits four excitation sources in the theoretical 3D diagnostic diagram. As shown in Figure~\ref{fig:3081_3d}, its spaxels populate a low-velocity-dispersion \ion{H}{2}$-$AGN mixing sequence, a nearly vertical \ion{H}{2}$-$pure-shock mixing sequence, and a distinct group of fast-shock-dominated spaxels. Figure~\ref{fig:3081_2d} presents zoomed-in 2D fractional maps of the central $\sim3\times3\,$kpc region. The \ion{H}{2}-dominated spaxels trace a ring-like structure composed of multiple star-forming clumps at a projected radius of $\sim1\,$kpc from the nucleus. AGN-photoionized emission is located primarily interior to this ring and dominates along the north-south axis, while also extending east-west, in contrast to the well-defined ionization cones seen in other galaxies in our sample. Fast-shock-dominated spaxels are confined to the central region ($r\lesssim300\,$pc) and extend along the east-west axis. Most pure-shock-dominated spaxels lie in the outer star-forming ring at $r\approx5\,$kpc, as shown in the full-field 2D fractional maps in Appendix~\ref{sec:appendixB} (Figure~\ref{fig:A1}).

In Figure~\ref{fig:3393_3d}, the NGC\,3393 spaxels show the characteristic distribution in the theoretical 3D diagram expected for galaxies with all four excitation sources, tracing an \ion{H}{2}$-$AGN mixing sequence, an \ion{H}{2}$-$pure shock mixing sequence, and an AGN$-$fast shock mixing sequence. The fractional 2D maps in Figure~\ref{fig:3393_2d} show that \ion{H}{2}-dominated emission in NGC\,3393 arises primarily from two star-forming arcs to the north and south, starting at a projected distance of $\sim3\,$kpc from the nucleus. AGN-dominated spaxels form two elongated cones (each $\sim3\,$kpc in length and $\sim4.6\,$kpc in width) extending from the nucleus towards the northeast and southwest directions. We also identify an AGN-photoionized arc located $\sim7.8\,$kpc northeast of the nucleus. Fast shock-dominated emission peaks in the central region and extends up to $1\,$kpc perpendicular to the AGN ionization axis. Pure-shock contributions are detected both surrounding the central fast-shock region and in the \ion{H}{2}-dominated arcs.  

Figure~\ref{fig:5643_3d} shows that NGC\,5643 exhibits the characteristic \ion{H}{2}–AGN mixing sequence and \ion{H}{2}–pure-shock mixing sequence in the theoretical 3D diagram, with a small subset of spaxels extending onto the fast-shock model grid. The 2D fractional maps in Figure~\ref{fig:5643_2d} further reveal that \ion{H}{2}-dominated emission is associated with several prominent star-forming structures: a large clump with a diameter of $\sim0.8\,$kpc located $\sim1.3\,$kpc east of the nucleus, an elongated star-forming region at a similar projected distance to the northwest, and more sparsely distributed clumps at larger radii ($\gtrsim2\,$kpc). AGN-dominated spaxels delineate a biconical structure extending to $\sim1.6\,$kpc from the nucleus along the east–west direction. Fast shock emission dominates near the nucleus and extends perpendicular to the AGN ionization cone axis. Pure shock-dominated spaxels are found both in the vicinity of the central fast-shock region and more commonly, near the outer star-forming clumps.

NGC\,7212 is the only galaxy in our sample that is adequately described by three excitation sources. As shown in the top panel of Figure~\ref{fig:7212_3d}, a small subset of spaxels lies near the \ion{H}{2} model grid, while the majority trace an AGN–fast-shock mixing sequence. No spaxels populate the pure-shock locus, possibly because pure-shock emission is diluted by neighboring AGN or \ion{H}{2} emission at the relatively coarse physical resolution (535 pc/$\arcsec$) of NGC\,7212. Figure~\ref{fig:7212_2d} presents zoomed-in 2D fractional maps of the central $\sim2\times2\,$kpc region, where \ion{H}{2}-dominated spaxels are largely absent. AGN-dominated emission is concentrated near the nucleus and extends up to $\sim2\,$kpc to the north and south, whereas fast shock-dominated emission forms an elongated structure perpendicular to the AGN ionization axis. The \ion{H}{2}-dominated spaxels are instead found primarily in a star-forming arc located $\sim6\,$kpc northeast of the nucleus, as shown in the full-field fractional maps in Appendix~\ref{sec:appendixB} (Figure~\ref{fig:A2}).

In summary, eight of the nine galaxies in our sample require all four excitation mechanisms, whereas NGC\,7212 is adequately described by three, with no pure-shock excitation required. Kiloparsec-scale star-forming rings or arcs and similarly extended AGN NLRs are present in all nine galaxies. Table~\ref{tab:2} summarizes the projected radii of the star-forming regions, the projected extents of the AGN NLRs, and the presence of shock excitation for each galaxy.

\section{Comparisons with X-ray observations}\label{sec:result}

We compare our excitation-source decomposition with Chandra X-ray imaging and spectroscopy to provide an independent validation of the inferred AGN- and shock-powered regions. Of the nine galaxies in our sample, eight have deep Chandra exposures ($t_{\rm exp}>100\,$ks) that enable the detection of kpc-scale extended soft X-ray emission and provide sufficient signal-to-noise for X-ray spectral analysis. The only exception is NGC\,424, for which only a shallow Chandra observation is currently available and thus we exclude it from the analysis below. 

Through uniform Chandra spectral modeling of 20 nearby heavily obscured AGNs, R. Middei et al. (2026, submitted; hereafter RM26) find that both circumnuclear and kpc-scale soft X-ray emission generally require a combination of multiple photoionized components and thermal (collisionally excited) components to reproduce the observed spectra. This result is qualitatively consistent with our optical decomposition, which indicates that AGN photoionization and shock excitation commonly coexist within the central kpc of these Seyfert galaxies. 

To assess the spatial correspondence between optical and X-ray diagnostics, we compare contours from X-ray bands dominated by photoionized emission to the AGN fractional-contribution maps, and contours from X-ray bands dominated by thermal emission to the combined shock (fast shock and pure shock) fractional-contribution maps in this section. The photoionization-dominated energy band is selected based on the best-fit X-ray spectral models of RM26, as summarized in Table~\ref{tab:2}. The thermal-dominated band is set to $0.9-1.2\,$keV, which includes Ne~IX, Ne~X, and Fe-L emission lines and has been linked to the presence of shocks because it traces radio lobe emission \citep{wang_2011_4151,paggi_2012_mrk573,krol_2026_eso137}. This energy range is also dominated by thermal emission in the X-ray spectral modeling of RM26. We note that a completely clean spatial separation of thermal and non-thermal emission is not possible, because both components generally contribute within the chosen bands, even when one dominates. Nevertheless, differences in X-ray contour morphologies are sufficiently pronounced in most cases to support our optical decomposed AGN- and shock-emission.

\begin{figure*}[hbt]
\epsscale{0.95}
\plotone{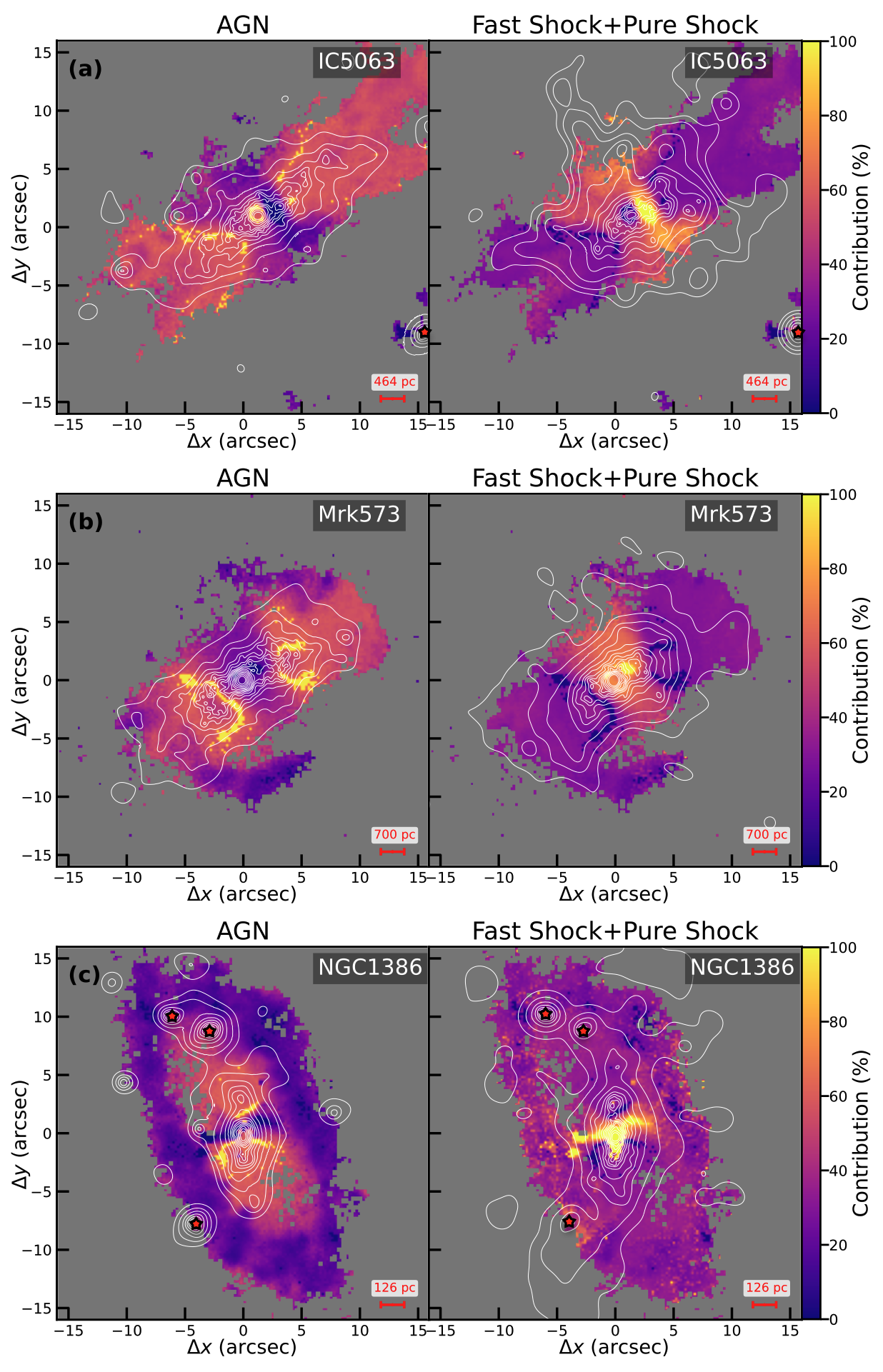}
\caption{Comparisons between Chandra X-ray contours and optically decomposed 2D fractional-contribution maps for IC\,5063, Mrk\,573, and NGC\,1386. Left panels: X-ray contours in the photoionization-dominated energy band (white) overlaid on the optical AGN fractional-contribution maps. Right panels: X-ray contours in the thermal-dominated energy band (white) overlaid on the optical shock (fast+pure shock) fractional-contribution maps. {\color{black}Red stars outlined in black mark X-ray point sources identified by RM26 as likely contaminants, such as X-ray binaries.}}
\label{fig:Xray1}
\end{figure*}

\begin{figure*}[hbt]
\epsscale{0.98}
\plotone{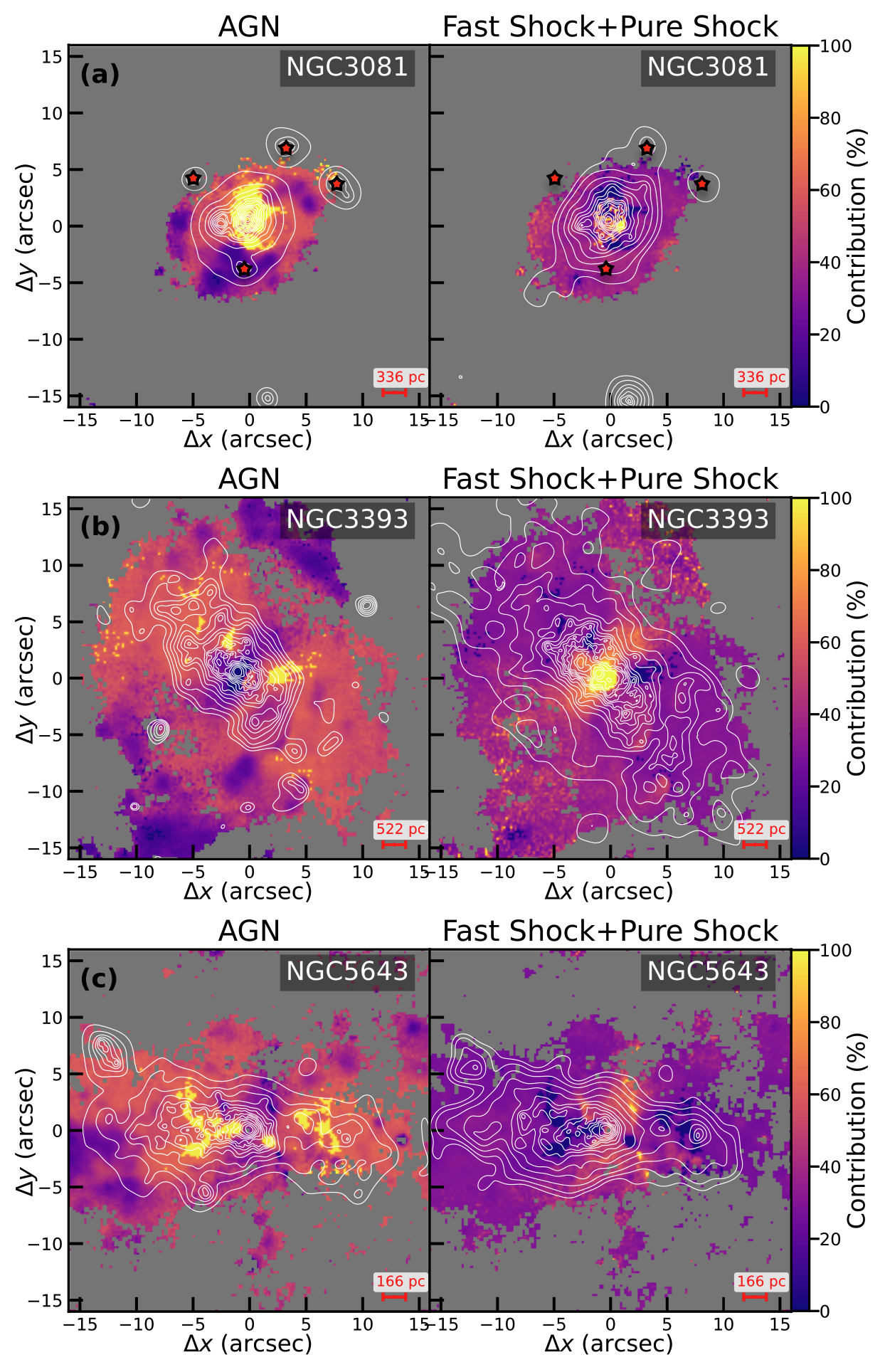}
\caption{Similar to Figure~\ref{fig:Xray1}, this figure compares Chandra X-ray contours with optically decomposed 2D fractional-contribution maps for NGC\,3081, NGC\,3393, and NGC\,5643. {\color{black}Red stars outlined in black mark X-ray point sources identified by RM26 as likely contaminants, such as X-ray binaries.}}
\label{fig:Xray2}
\end{figure*}

\begin{figure*}[hbt]
\epsscale{1.0}
\plotone{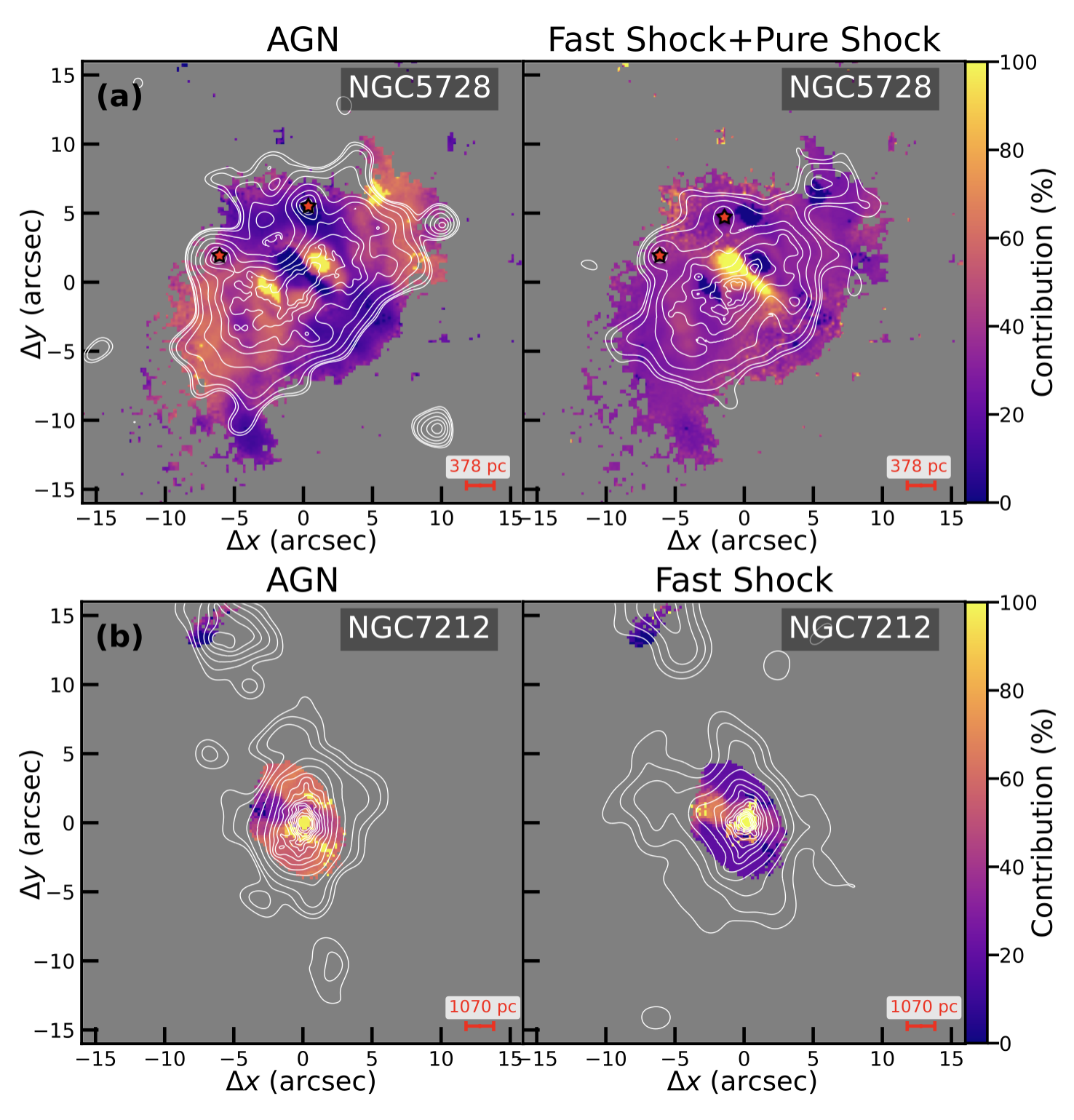}
\caption{Similar to Figure~\ref{fig:Xray1}, this figure compares Chandra X-ray contours with optically decomposed 2D fractional-contribution maps for NGC\,5728 and NGC\,7212. {\color{black}Red stars outlined in black mark X-ray point sources identified by RM26 as likely contaminants, such as X-ray binaries.}}
\label{fig:Xray3}
\end{figure*}

{\color{black}Some galaxies also show off-nuclear compact or point-like X-ray sources, likely contaminants such as X-ray binaries. These sources were excluded from the X-ray spectral fitting in RM26 and therefore do not affect the energy-band identification. However, they remain visible in the X-ray contour maps and can affect the visual comparison with the optical decomposition maps. We therefore mark the point sources identified by RM26 with red stars outlined in black in Figures~\ref{fig:Xray1}--\ref{fig:Xray3}. These point sources generally overlap with \ion{H}{2}-dominated regions, consistent with an origin related to compact stellar sources such as X-ray binaries.}

\setlength{\tabcolsep}{5pt}
\begin{deluxetable*}{c|c|c|c|c|c|c|c}[hbt]
\centering
\tablewidth{2pc}
\tablecaption{Summary of Excitation Sources and X-ray--Optical Correspondence for the Seyfert 2 Galaxy Sample\label{tab:2}}
\tablenum{2}
\tablehead{
  \colhead{Galaxy} &
  \colhead{SF Regions} &
  \colhead{AGN NLR} &
  \colhead{Fast Shocks} &
  \colhead{Pure Shocks} &
  \colhead{$E_{\rm photo}$} &
  \multicolumn{2}{|c}{X-ray--Optical Correspondence}
  \\
  \hline
  \colhead{} &
  \colhead{radius\tablenotemark{\scriptsize a}} &
  \colhead{(length, width)\tablenotemark{\scriptsize b}} &
  \colhead{Present?} &
  \colhead{Present?} &
  \colhead{(keV)\tablenotemark{\scriptsize c}} &
  \multicolumn{1}{|c}{Photoionized--AGN?} &
  \colhead{Thermal--Shocks?}
  }
\startdata
IC\,5063 & $\sim$4\,kpc & (3.5-5.3\,kpc, 2.3\,kpc) & Yes & Yes &$1.5-3.0$ & Yes & Yes  \\
MRK\,573 & $\sim$3\,kpc & (3\,kpc, 3.5\,kpc) & Yes & Yes &$0.3-0.8$ & Yes & Yes \\
NGC\,424\tablenotemark{\scriptsize d} & $\sim$4.8\,kpc & (1\,kpc, 1\,kpc) & Yes & Yes & -- & -- & -- \\
NGC\,1386 & $\sim$800\,pc & (600\,pc, 500\,pc) & Yes & Yes &$0.3-0.9$ & Yes & Yes \\
NGC\,3081 & $\sim$1\,kpc & $r\lesssim1\,$kpc & Yes & Yes &$2.0-4.0$ & Yes & Yes \\
NGC\,3393 & $\sim$3\,kpc & (3\,kpc, 4.6\,kpc) & Yes & Yes &$2.0-4.0$ & Yes & Yes\\
NGC\,5643 & $\sim$1.3\,kpc & (1.6\,kpc, 830\,pc) & Yes & Yes &$1.0-4.0$ & Yes & Yes \\
NGC\,5728 & $\sim$1\,kpc & (2\,kpc, 1\,kpc) & Yes & Yes &$1.5-3.0$ & Yes & Yes \\
NGC\,7212 & $\sim$6\,kpc & (2\,kpc, 4\,kpc) & Yes & No & $1.5-3.0$ & Yes & Yes \\
\enddata
\tablenotetext{\scriptsize a}{Projected radius of the star-forming ring or arc, measured from the galaxy center in the plane of the sky.}
\tablenotetext{\scriptsize b}{Projected length and maximum width of each AGN narrow-line region cone, measured in the plane of the sky. These values are limited by the observational depth of the MUSE data cube and may therefore underestimate the full spatial extent of the AGN NLRs.}
\tablenotetext{\scriptsize c}{Chandra energy band used to trace the photoionized X-ray component.}
\tablenotetext{\scriptsize d}{NGC\,424 is excluded from the X-ray--optical comparison because only a shallow Chandra observation is currently available.}
\end{deluxetable*}
\vspace{-2.em} 

The X-ray contours in IC\,5063 strongly support our optical decomposition of AGN and the distribution of shocks. As shown in Figure~\ref{fig:Xray1}a, the photoionized X-ray emission extends to $\gtrsim14\arcsec$ along the AGN bicone but remains confined along the cross-cones ($r\lesssim5\arcsec$), independently supporting the AGN ionization cones identified from the optical decomposition. The thermal component is slightly less extended along the bicone yet much broader in the cross-cone direction, reaching $r\gtrsim15\arcsec$ to the northeast. X-ray spectral fits show that, within the thermal energy band, the thermal component is comparable to or fainter than the photoionized component in the bicone but dominates in the cross-cone (RM26, see also Figures 17 and 19 in \citet{2021ApJ...921..129T}). The resulting thermal-band X-ray contours therefore primarily trace shock-heated gas in the cross-cone direction, providing an independent X-ray confirmation of the strong cross-cone shock contribution inferred from the optical decomposition.

The X-ray data are broadly consistent with our AGN–shock decomposition in Mrk\,573. In Figure~\ref{fig:Xray1}b, the photoionized-band X-ray contours are elongated along the AGN ionization bicone, coincident with the optically classified AGN-dominated regions. The thermal-band X-ray contours show excess emission in the cross-cone direction, where the optical decomposition identifies shocks. Similar to IC\,5063, the comparable bicone extension in the thermal- and photoionized-band contours can be explained by the X-ray spectral fits, which show that photoionized emission still contributes significantly to the bicone spectra even within the nominally thermal band in Mrk\,573 (RM26).

In NGC\,1386, the Chandra morphology likewise supports our optical decomposition. As shown in Figure~\ref{fig:Xray1}c, the photoionized-band contours closely follow the optically identified AGN-dominated cones, reinforcing the optical AGN decomposition. The thermal-band emission also extends within the bicone, suggesting that thermal plasma may contribute in this region, plausibly associated with interactions between the AGN outflow and the surrounding ISM. Independently, our optical decomposition indicates a $\sim40\%$ shock contribution in the bicone. In the cross-cone direction, the thermal-band contours extend to larger radii and spatially coincide with the optically identified shock-dominated regions perpendicular to the ionization cones.

In NGC\,3081, the optical decomposition shows that AGN photoionization dominates within the central $r\sim5\arcsec$. This is consistent with the morphology of the Chandra photoionized-band contours in Figure~\ref{fig:Xray2}a, which are concentrated in the same central region. The X-ray thermal-band contours overlap with the optically identified shock-dominated regions in the central $r\lesssim2\arcsec$, providing an independent morphological support for the optical decomposition. At larger radii ($r\gtrsim2\arcsec$), the thermal- and photoionized-band contours have comparable extents. Based on the RM26 X-ray spectral fits, this similarity likely reflects the fact that photoionized emission contributes substantially within the nominally thermal energy band.

NGC\,3393 and NGC\,5643 exhibit an X-ray–optical correspondence similar to that of other galaxies with biconical AGN structures. In Figures~\ref{fig:Xray2}b and c, the photoionized-band X-ray contours are elongated and aligned with the optically identified AGN ionization cones. The thermal-band contours show excess emission in the cross-cone direction, consistent with the optically inferred shocks that extend perpendicular to the bicone. X-ray spectral fits (RM26) indicate that photoionized emission contributes substantially to the flux in the adopted thermal band within the AGN bicones. Therefore, thermal-band contours also extend along the bicone even if shocks are not the dominant contributor there. 

In NGC\,5728, the Chandra morphology also supports our optical decomposition. As shown in Figure~\ref{fig:Xray3}a, the photoionized-band X-ray contours closely trace the optically identified AGN-dominated bicone, while the thermal-band contours show a more extended distribution in the cross-cone direction, consistent with the central shock-dominated region inferred from the optical decomposition. Further support comes from an optically identified shock-dominated, elongated feature ($>50\%$) located $\sim5\arcsec$ east of the nucleus, whose morphology closely matches the thermal-band contours. 

NGC\,7212 shows a clear morphological difference between the photoionized- and thermal-band X-ray emission (Figure~\ref{fig:Xray3}b). The photoionized-band contours are elongated along the AGN ionization axis and remain relatively confined in the cross-cone direction, whereas the thermal-band contours are broader transverse to the cones. This cross-cone thermal emission coincides with the optically identified shock-dominated regions, consistent with the expectation that shock-heated gas contributes excess soft X-ray emission. However, the limited X-ray photon counts in the cross-cone regions prevent a robust spectral test of whether an additional thermal component is required in the RM26 spectral fits.

As summarized in Table~\ref{tab:2}, the Chandra X-ray morphology provides independent, qualitative support for our optical decomposition in the eight galaxies with available X-ray coverage. The photoionized-band contours are broadly aligned with the optically identified AGN-dominated regions, while the thermal-band contours often show extended soft X-ray emission in the cross-cone direction, broadly coincident with regions where the optical decomposition identifies enhanced shock contributions. This is consistent with a scenario in which soft X-ray emission along the ionization bicone is dominated by photoionized gas, whereas thermal emission in the cross-cone direction traces shock-heated gas associated with interactions between AGN-driven outflows and the surrounding ISM. Similar behavior has been reported in high-resolution Chandra observations of nearby Seyfert galaxies \citep[e.g.][]{falcao_deep_2024}.

\section{Discussion}\label{sec:discussion}

\subsection{Common Patterns of AGN, SF, and Shocks in Seyfert Galaxies}

Our decomposition reveals a recurring and coherent spatial pattern across the Seyfert galaxies in our sample. Star-formation–dominated spaxels primarily trace circumnuclear rings and arcs on kpc scales, while AGN-dominated spaxels delineate ionization bicones extending from the nucleus. In addition to these two commonly identified components, we find that fast shocks are concentrated in the central regions ($\lesssim$500 pc) and often extend preferentially in the cross-cone direction, perpendicular to the AGN ionization axis. In most galaxies, pure shocks appear both around the central fast-shock region and more sparsely within star-forming rings and arcs, suggesting that shocks can arise from both AGN-driven and stellar-driven outflows, depending on location. A similar coexistence of ionization cones, circumnuclear star-forming rings, and shock-excited gas has also been reported in nearby AGNs based on deep Chandra X-ray and HST narrow-band imaging \citep[e.g.][]{falcao_deep_2024,falcao_ngc5005_2025,krol_eso137_2025}.

The coexistence of circumnuclear star-forming structures and AGN-ionized cones, often interpreted as evidence for ongoing AGN feedback, has been reported for several galaxies in our sample (IC\,5063, NGC\,1386, NGC\,5643, and NGC\,5728) using standard 2D optical diagnostic diagrams \citep{cresci_magnum_2015,mingozzi_magnum_2019,durre_agn_2018,shin_positive_2019}. For instance, in NGC\,5643 the star-forming regions located near the ionization cones have been discussed as possible signatures of positive AGN feedback \citep{cresci_magnum_2015}. For NGC\,5728, \citet{shin_positive_2019} analyzed the spatial distribution of star-formation efficiency on the circumnuclear ring and reported evidence for both positive and negative feedback operating in different ring segments. Nevertheless, this interpretation of the feedback is limited by geometric ambiguity: it is often unclear whether the star-forming structures and AGN-ionized gas are physically interacting in three dimensions or simply overlapping in projection on the sky.

Several studies have identified enhanced velocity dispersion and disturbed kinematics in the inner kpcs in Seyfert galaxies, sometimes with extensions transverse to the AGN ionization axis \citep[e.g.][]{couto_kinematics_2013,riffel_outflow_2014,finlez_jet_2018,mingozzi_magnum_2019,venturi_magnum_2021,zhang_galaxy_2024}. In the 2D [SII]/H$\alpha$-[OIII]/H$\beta$ diagram, \citet{venturi_magnum_2021} found these high-$\sigma$ spaxels typically fall in the LINER-like regime. However, their dominant excitation mechanism remains uncertain: LINER-like line ratios can arise from AGN photoionization \citep[e.g.][]{ho_nuclear_2008}, shocks \citep[e.g.][]{dopita_probing_2015}, or other processes \citep[e.g.][]{belfiore_sdss_2016}, and the degeneracy cannot be resolved with 2D line-ratio diagnostics alone. 

A key result of this work is that we show, for the first time, that the high-$\sigma$ regions highlighted in previous works are consistent with shock excitation in both their optical emission line ratios and kinematics. In our sample, elevated-$\sigma$ spaxels systematically occupy the shock model loci in the theoretical 3D diagnostic diagram and form mixing sequences connecting to either the HII or AGN model grid, indicating star formation–shock and AGN–shock mixing, respectively. Notably, although the 3D diagram contains no spatial information, the shock-dominated spaxels assemble into coherent structures in the 2D maps, revealing physically meaningful morphologies in these Seyfert galaxies.

In addition to the X-ray correspondence discussed in Section~\ref{sec:result}, near-infrared coronal-line studies also reveal evidence of shocks in several galaxies in our sample. In NGC\,1386, \citet{rodriguezardila_ngc1386_2017} detected central [\ion{Si}{6}]~1.963\,$\mu$m emission within $\lesssim1\arcsec$, indicative of shocks with $v_s\gtrsim200\,{\rm km\,s^{-1}}$. More recently, \citet{rodriguezardila_CLR_2025} studied coronal-line regions in nine Seyfert galaxies, including six objects in our sample: IC\,5063, NGC\,1386, NGC\,3081, NGC\,3393, NGC\,5643, and NGC\,5728. They found that the radial decline of coronal-line fluxes within $r\lesssim1$\,kpc is shallower than expected for ionization by a central source alone, as expected in the presence of central shocks. The presence of shocks is further supported by their finding that extended coronal emission accounts for more than 50\% of the total coronal emission in some galaxies, with high-ionization coronal lines appearing off-nucleus.

Complementary infrared IFU observations further suggest that these shocks arise in a stratified multiphase ISM. In IC\,5063, JWST/MIRI observations revealed ISM stratification in the high-$\sigma$ regions, with the ionized gas exhibiting larger velocity dispersion than the molecular gas \citep{dasyra_case_2024}. Enhanced [\ion{Fe}{2}]/Pa$\alpha$ and H$_2$~S(1)/Pa$\alpha$ ratios were also found in the same regions by \citet{dasyra_radio_2015} using VLT/SINFONI data, as expected in the presence of shocks \citep{mouri_FeII_2000}. A similar multiphase structure emerges in NGC\,5728, where \citet{davies_gatos_2024} used JWST/MIRI observations to study the excitation of H$_2$ gas and identified slow shocks with $v_s\approx30\,{\rm km\,s^{-1}}$ in dense molecular gas ($n\sim10^5\,{\rm cm^{-3}}$). These slow molecular shocks are spatially coincident with the region where we identify strong fast shocks in the ionized gas. Similarly, \citet{delaney_agn_2025} found evidence for slow-shock excitation of H$_2$ gas in the nuclear region of NGC\,3081. Taken together, these results point to a stratified multiphase ISM, in which ionized gas closer to the nucleus is affected by faster shocks and larger velocity dispersions, while more extended molecular gas is associated with slower shocks and lower velocity dispersions.

The emergence of a common pattern across star-forming regions, AGN narrow-line regions, and shock-excited gas makes understanding its physical origin a key next step. In the following subsections, we discuss our current interpretation of this behavior in the context of existing theoretical and observational work.

\subsection{How do Circumnuclear Star-forming Rings form?}

The ubiquity of circumnuclear star-forming rings across our sample, together with their coexistence with AGN activity and central shocks, provides a unique opportunity to study their formation mechanism. Two scenarios are particularly relevant: (1) the bar-driven inflow scenario, in which non-axisymmetric torques drive gas inward, allowing it to accumulate near dynamical resonances and form circumnuclear star-forming rings; and (2) the AGN positive-feedback scenario, in which shocks driven by jet- or wind-ISM interactions compress the gas and enhance star formation in ring-like structures. Below, we discuss how our results relate to these two mechanisms. 

\subsubsection{Bar-driven Gas Inflow Scenario}

Based on the RC3 classifications from optical B-band imaging \citep{1991rc3..book.....D}, seven of the nine galaxies in our sample host bars. These barred systems also exhibit prominent star-forming rings or arcs in our optical decomposition, with projected radii of $r\sim0.8-4\,$kpc. This qualitative correspondence between large-scale bars and circumnuclear star-forming rings is consistent with the bar-driven inflow scenario, although it does not rule out a possible contribution from AGN positive feedback.

In this picture, the circumnuclear star-forming ring forms as a consequence of bar-driven gas inflow. Non-axisymmetric torques induced by the bar remove angular momentum from the gas and funnel it inward, where it can stall and accumulate near Lindblad resonances. The resulting dense gas concentrations can then fuel circumnuclear star formation \citep{1992MNRAS.259..345A,buta_rings_1996,kormendy_secular_2004}. 

However, circumnuclear rings can also form in the absence of large-scale bars. In our sample, IC\,5063 and NGC\,7212 do not show clear large-scale bar structures in the RC3 classification, yet both host circumnuclear star-forming structures. In addition, mid-infrared morphological analysis finds no evidence for a bar in NGC\,1386 \citep{buta_morphology_2015}, suggesting that the optical classification may be affected by dust obscuration. Circumnuclear rings in non-barred galaxies have also been reported by \citet{comeron_rings_2010}, who found that 18 of 107 galaxies with nuclear rings are classified as non-barred. Therefore, star-forming rings are not restricted to barred systems.

One plausible explanation for rings in apparently non-barred systems is that resonances can also be driven by weaker non-axisymmetric perturbations, such as oval distortions, spiral structure, tidal interactions, or minor mergers \citep{knapen_unbarred_2004,comeron_rings_2010}. Testing whether the ring radii observed in our sample are consistent with resonance expectations will require detailed dynamical modeling \citep[see, e.g.,][]{mazzuca_ring_2011}.

The same inflow process that builds circumnuclear star-forming rings may also help transport gas further inward to fuel the central black hole. High-resolution simulations of gas inflow in Seyfert and barred disk galaxies predict that gravitational instabilities on scales of $\lesssim 500$ pc can generate nested circumnuclear morphologies, including nuclear spirals, rings, nuclear bars, barred rings, clumpy discs, and gas-rich streams \citep[e.g.,][]{sholosman_bars_1989,englmaier_nuclear_2000,
hopkins_gas_2010,angles-alcazar_2021}. These non-axisymmetric structures can efficiently remove angular momentum efficiently and channel gas inward from hundreds of parsecs to parsec scales, thereby sustaining nuclear activity \citep{sholosman_agn_1990}. In NGC\,3081, a nuclear bar has been reported within the central $r\sim2\arcsec$ by \citet{schnorr-muller_feeding_2016}, consistent with this general picture. However, the nested ``ring-within-ring'' and ``bar-within-bar'' structures predicted by these models cannot yet be robustly assessed with the current data. Higher-spatial-resolution observations will be needed to determine whether such multi-stage inflow is present in these galaxies.

\subsubsection{Positive Feedback from AGN Jets or Winds Scenario}\label{sec:discussion22}

In the AGN positive-feedback scenario, interactions between AGN jets or winds (outflows) and the surrounding ISM can drive shocks that compress the gas, increase its density, and enhance local star formation. In jet-ISM simulations, this enhancement can occur in ring-like regions close to the jet axis on scales of $r\lesssim1\,$kpc \citep{mukherjee_relativistic_2018}. At later times, the expanding jet cocoon can also pressurize the disk ISM more globally, producing enhanced star formation at larger radii of $r\sim3-6\,$kpc \citep{gaibler_jet_2012,dugan_jet_wind_2017}. Similar large-scale positive feedback at $r\gtrsim3\,$kpc has also been found in simulations that include wide-angle AGN winds without jets \citep{dugan_jet_wind_2017}.

The coexistence of circumnuclear star-forming rings, AGN activity, and central shocks across our sample therefore makes positive AGN feedback a plausible contributor to the formation of these rings. The projected radii of the star-forming rings and arcs in our sample, ranging from $r\sim\, 1$ kpc in NGC\,1386, NGC\,3081, NGC\,5643, and NGC\,5728 to $r\sim3-6\,$ kpc in other galaxies, are broadly consistent with the scales over which positive feedback is predicted to operate in these simulations. However, this spatial agreement alone does not uniquely distinguish AGN-triggered star formation from bar-driven inflow or other dynamical mechanisms. 

The key distinguishing evidence lies in the stellar kinematics of the star-forming rings. Although positive feedback hydrodynamic simulations differ in their predicted star-formation enhancement factors, delay times after jet launching, and durations of the enhanced star-formation phase \citep[e.g.][]{gaibler_jet_2012,dugan_jet_wind_2017,mukherjee_relativistic_2018}, they share a common expectation: stars formed through positive feedback should retain kinematic signatures of the jet- or wind-driven disturbance. In particular, these stars are expected to exhibit high outward radial velocities that deviate from regular disk rotation. In contrast, stars formed in bar-driven inflow rings should primarily follow regular circular rotation, possibly with only modest inward radial motions.

Testing whether the circumnuclear star-forming rings in our sample result from positive feedback will therefore require detailed analysis of stellar absorption features within the rings using higher-spectral-resolution data. Such analysis is beyond the scope of the present paper, but it will be important for distinguishing AGN positive feedback from bar-driven inflow as the origin of the circumnuclear star-forming rings.

\subsection{What Generates the Central Shocks?}\label{sec:discussion3}

Since the discovery of enhanced velocity-dispersion regions, which we now confirm to be excited by shocks, in the central regions of Seyfert galaxies, AGN jet–ISM interaction has been the leading explanation for their origin \citep[e.g.,][]{mukherjee_relativistic_2016,venturi_magnum_2021}. However, AGN wind-ISM interactions may also play an important role, particularly in galaxies hosting low-power jets \citep[e.g.][]{ciotti_feedback_2017,yuan_active_2018}, and can trigger central shocks when coupled with a weak jet \citep{guo_coupling_2026}. In the following subsections, we examine how our results relate to these two mechanisms.

\subsubsection{Jet-ISM Interactions}

In the jet-ISM interaction scenario, a powerful jet with jet power of $P_{jet}\geq10^{43}\,\rm erg\, s^{-1}$ propagating through a dense, turbulent, two-phase ISM can inflate energy bubbles that perturb the surrounding gas on kiloparsec scales \citep{mukherjee_relativistic_2016}. As these bubbles expand, shocks form at their boundaries through interactions with the ambient ISM, naturally producing a multiphase medium and driving radial outflows \citep{mukherjee_relativistic_2016}. When the jet is launched into a gaseous disk, shocks driven by both the jet and the expanding energy bubble can raise the velocity dispersion of the ionized gas ($T>10^4\,$K) throughout the disk to $\sigma\sim400-800\,$km/s, while the colder gas ($T<10^4\,$K) remains $\sigma\lesssim100\,$km/s \citep{mukherjee_relativistic_2018}. Similar values are also obtained when the jets are inclined towards the disk.

Our results are consistent with the jet-ISM interaction scenario in five respects: 
(1) all Seyfert galaxies in our sample show radio-detected jets extending broadly along the AGN ionization direction\citep{ulvestad_radio_1984,schommer_ionized_1988,falcke_VLA_1998,morganti_radio_1999,nagar_VLA_1999,leipski_radio_2006,mundell_radio_2009}; 
(2) with the exception of NGC\,1386 and NGC\,3081 (discussed below), the remaining galaxies have empirically inferred jet powers of $P_{\rm jet}\sim10^{42}$--$10^{44}\,{\rm erg\,s^{-1}}$ (D.~Krol et al., 2026, in prep.), estimated from the 1.4\,GHz NVSS flux densities reported in NED using the $P_{\rm jet}$--$P_{\rm 1.4\,GHz}$ scaling relation of \citet{cavagnolo_radio_2010}; 
(3) the high-$\sigma$ regions are elongated perpendicular to the jets, consistent with the morphology predicted by jet-ISM simulations; 
(4) their optical line ratios are consistent with radiative shock model predictions; and
(5) a stratified multiphase ISM is observed in IC\,5063, NGC\,5728, and NGC\,3081, with high-$\sigma$ warm ionized gas spatially coincident with cooler, lower-$\sigma$ gas ($\sigma\lesssim100\,$km/s).

Indeed, \citet{mukherjee_jet-ism_2018} conducted a case study of IC\,5063 and showed that jet-ISM interaction simulations with jet power of $10^{44}-10^{45}\,\rm erg\, s^{-1}$ can reproduce the disturbed kinematics observed across the ionized, neutral, and molecular gas phases. The jet power we infer for IC\,5063, $P_{jet}\approx(7.0\pm2.9)\times10^{43}\,\rm erg\, s^{-1}$, is about an order of magnitude lower than the value adopted in those simulations. This difference is not unexpected, because our estimate is based on an empirical scaling relation between radio luminosity and cavity power calibrated for evolved radio jets propagating through smooth halos of galaxies or clusters. Such relations may underestimate the instantaneous mechanical power of young or compact jets interacting with the clumpy ISM of a gaseous galactic disk, as in IC\,5063 \citep{mukherjee_jet-ism_2018}. Under this interpretation, simulations exploring the $P_{jet}\geq10^{43}\,\rm erg\, s^{-1}$ regime \citep{mukherjee_relativistic_2016} are broadly applicable to galaxies with observed jet powers of $P_{jet}\geq10^{42}\,\rm erg\, s^{-1}$, which includes the majority of our sample.

Independent X-ray studies provide further support for the jet-ISM interaction scenario. Extended soft X-ray emission oriented perpendicular to the radio jet (in the cross-jet direction) has been reported in several nearby Compton-thick AGNs observed with Chandra, including ESO 428-G014 \citep{fabbiano_jet_2018}, IC\,5063 \citep{2021ApJ...921..129T}, NGC\,1167 \citep{fabbiano_jet-ism_2022}, and NGC\,5728 \citep{falcao_deep_2023}. Such cross-jet X-ray structures are naturally reproduced in hydrodynamic jet–ISM simulations when X-ray emissivities are computed self-consistently, where shocked gas produces strong thermal X-ray emission both near the jet head and in regions lateral to the jet axis \citep{fabbiano_jet-ism_2022}. 

The jet–ISM interaction scenario also makes additional testable predictions on sub-kpc scales \citep{mukherjee_relativistic_2018}, including inflows within the gaseous disk driven by backflows from jet-inflated energy bubbles, as well as enhanced star formation in a ring-like geometry induced by shock compression (discussed in Section~\ref{sec:discussion22}). Testing these predictions will require higher-spatial-resolution observations across multiple wavelengths to resolve the relevant gas and stellar motions in greater detail.

Jet–ISM interactions, although appealing, may not provide the full explanation in every case. Based on the empirical scaling relation from \citet{cavagnolo_radio_2010}, two galaxies in our sample, NGC\,1386 ($P_{jet}\approx4.6\times10^{41}\,\rm erg\,s^{-1}$) and NGC\,3081 ($P_{jet}\approx5.1\times10^{41}\,\rm erg\,s^{-1}$) appear to host relatively low-power jets. The jet–ISM simulations discussed above are only applicable to $P_{jet}\geq10^{42}\,\rm erg\, s^{-1}$ systems, and recent simulations covering the $P_{jet}=10^{38}-10^{43}\,\rm erg\, s^{-1}$ regime suggest that low-power jets drive warm ionized outflows only under specific conditions \citep{borodina_jet_2025}. Further studies of low-power jets launched into gaseous disks are needed to determine whether they can account for the central shocks and the elevated$-\sigma$ gas observed perpendicular to the jet axis in NGC\,1386 and NGC\,3081.

\subsubsection{AGN Wind-ISM Interactions}

In radiative-mode AGNs, such as the Seyferts in our sample, the mechanical feedback from AGN-driven winds can inject an amount of energy comparable to radiative feedback, while also playing a dominant role in suppressing black hole accretion \citep{guo_feedback_2014,ciotti_feedback_2017,yuan_active_2018,2026ApJ..1000...41Z}. AGN winds have also been proposed as a mechanism for producing X-ray cavities in Seyfert galaxies \citep{yuan_active_2018}, and recent simulations suggest that winds coupled with weak jets can generate central shocks \citep{guo_coupling_2026}. This raises the possibility that, even in galaxies without prominent jets, or in systems hosting only low-power jets ($P_{jet}\lesssim10^{42}\,\rm erg\,s^{-1}$), such as NGC\,1386 and NGC\,3081, interactions between AGN-driven winds and the ambient ISM may still contribute to the observed central fast shocks and to the overall mechanical-energy budget.

Support for this picture may come from the population of so-called red geysers identified in the MaNGA survey \citep{cheung_suppressing_2016,roy_detecting_2018}. These galaxies are quiescent, predominantly early-type galaxies that exhibit bisymmetric elevated-EW(H$\alpha$) structures interpreted as signatures of AGN-driven winds \citep{roy_evidence_2021}. \citet{roy_radio_2021} studied the radio properties of 140 local early-type red geyser galaxies and found that only 42 are radio detected, with inferred jet powers mostly in the regime $P_{jet}\lesssim10^{43}\,\rm erg\,s^{-1}$. These galaxies also show elevated central velocity dispersions, in some cases extending perpendicular to the bisymmetric elevated-EW(H$\alpha$) features. Interestingly, two red geysers in their sample display extended radio jets oriented perpendicular to the AGN-wind axis: in one case the elevated-$\sigma$ gas extends perpendicular to the jet, while in the other it extends along the jet and thus perpendicular to the AGN-wind axis. Further studies that disentangle the relative roles of jet–ISM and wind–ISM interactions in this population will provide important insight into the nature of AGN-driven winds and their impact on the host galaxy.

Nevertheless, current AGN wind-ISM interaction simulations have not yet established whether AGN winds can reproduce the full range of velocity dispersions and spatial morphologies of the high-$\sigma$ regions observed in our Seyfert sample. It also remains unclear whether AGN wind-ISM interactions alone can generate central shocks, or whether coupling with a weak jet is required. Future simulations tailored to Seyfert-like disk galaxies, combined with spatially resolved comparisons between ionized-gas kinematics and X-ray morphology, will be necessary to quantify the relative roles of jets and winds.

\section{Conclusion}\label{sec:conclusion}

This paper investigates the spatial distribution of star formation, AGN photoionization, and shock excitation in nine Type 2 Seyfert galaxies using VLT/MUSE IFU observations. We apply a recently developed theoretical 3D diagnostic diagram to separate the contributions from star formation, AGN photoionization, fast shocks, and pure shocks. Most strikingly, all nine galaxies exhibit a common excitation structure:
\vspace{-0.5em}
\begin{itemize}
    \item Star-formation-dominated regions form ring-like or arc-like structures at projected radii of $r\sim0.8$--$6\,{\rm kpc}$ from the nucleus.
    \item AGN narrow-line regions exhibit biconical morphologies and extend to projected distances of $\sim1$--$5\,{\rm kpc}$ from the galaxy center.
    \item Fast-shock (shock plus precursor) dominated regions are consistently present near the galaxy center and often extend up to $\sim1\,{\rm kpc}$ perpendicular to the AGN ionization axis.
    \item Pure-shock (shock-only) dominated regions appear in two spatial groups: one surrounding the central fast-shock regions and another located within the star-forming rings or arcs. The former likely reflects ISM inhomogeneity, which can suppress or weaken the precursor contribution, while the latter is likely associated with stellar feedback.
\end{itemize}
\vspace{-0.5em}
Comparison with deep Chandra X-ray observations provides independent support for our optical decomposition in two ways:
\begin{itemize}
    \item The X-ray photoionized-band contours spatially agree with the optically decomposed AGN-dominated regions.    
    \item The X-ray thermal-band contours often show excess emission in the cross-cone direction, consistent with the extended shock features perpendicular to the AGN bicone identified in our optical decomposition.
\end{itemize}
\vspace{-0.5em}

We also discuss possible mechanisms responsible for the circumnuclear star-forming rings and central shocks in these Seyfert galaxies. The star-forming rings may be produced by bar-driven gas inflow, AGN positive feedback, or a combination of both. The central shocks may arise from jet--ISM interactions, or from wind--ISM interactions aided by weak jets in systems with low-power radio jets.

Overall, the identification of central shocks in our Seyfert sample provides a powerful and previously underexploited constraint on AGN mechanical-energy injection. By combining measurements of ISM pressure, gas density, and gas-phase metallicity in shock-dominated regions, it becomes possible to estimate the energy carried by the shocked gas and to trace how feedback energy is deposited and transported through the circumnuclear ISM (P.~Zhu et al., 2026b, submitted). Together with the excitation-source decomposition developed in this work, these measurements will provide an independent way to quantify how AGN mechanical feedback couples to, energizes, and redistributes gas in the circumnuclear ISM.

\section{Acknowledgement}
P.Z. would like to thank Yuan Fang, Yuxuan Zou, and Minhang Guo for useful discussions on AGN feedback mechanisms. ATF was supported by an appointment to the NASA Postdoctoral Program at the NASA Goddard Space Flight Center, administered by Oak Ridge Associated Universities under contract with NASA. RM acknowledges financial support from the INAF Scientific Directorate. {\color{black}This work was in part supported by the Smithsonian Institution's Combined Call for Research Award ``Fingerprints of Black Hole Feedback: X-ray Answers to a CosmicMistery'' (PI: G. Fabbiano; co-PI M. Elvis).}

\bibliography{Paper_Sep.bib}{}
\bibliographystyle{aasjournal}

\appendix
\renewcommand{\thefigure}{A\arabic{figure}}
\setcounter{figure}{0}
\section{Theoretical 3D diagrams} \label{sec:appendixA}
\begin{figure*}[hbt]
\begin{interactive}{js}{Mrk573ppxf_int.zip}
\epsscale{1.13}
\plotone{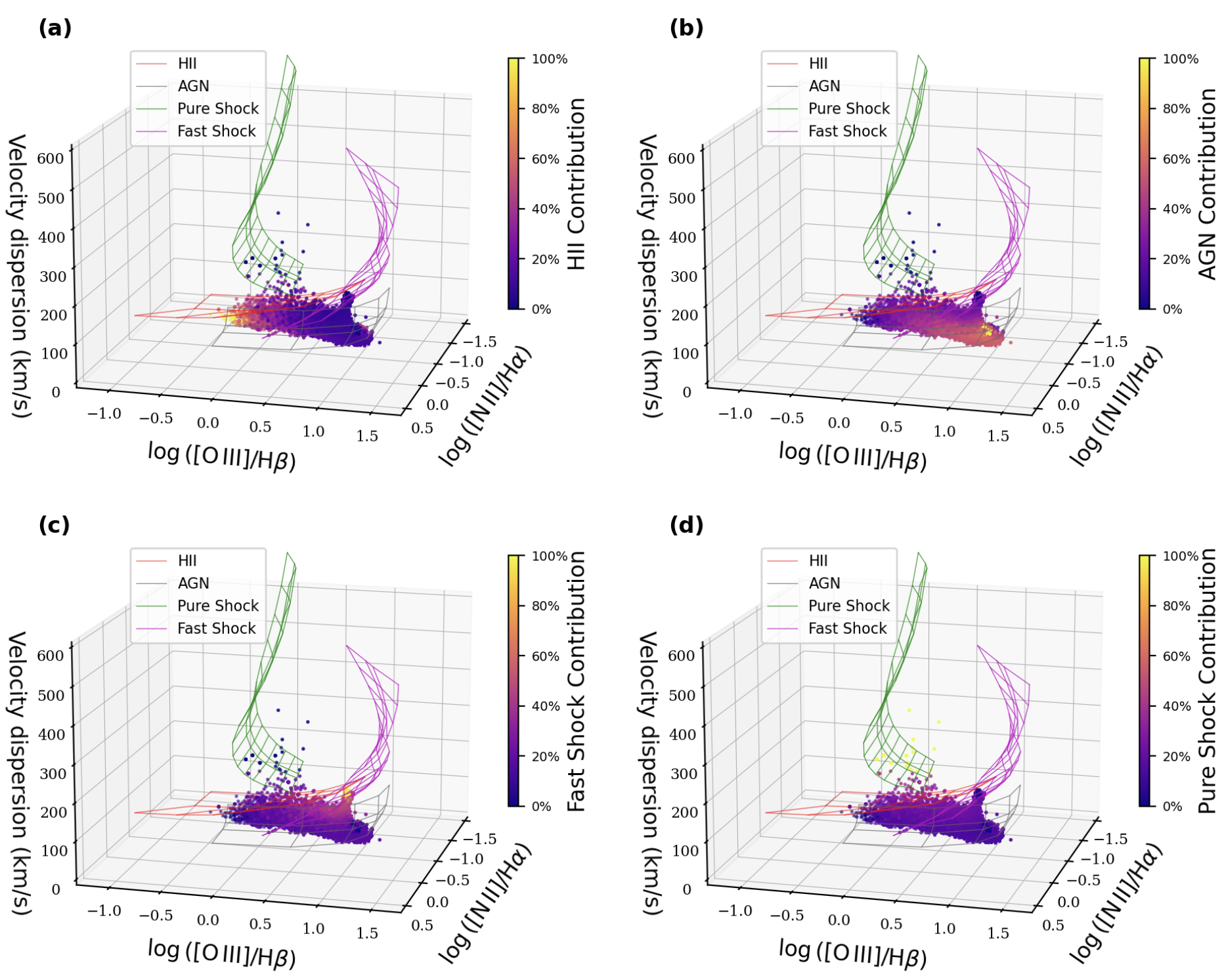}
\end{interactive}
\caption{Theoretical 3D diagrams of the MUSE IFU data for Mrk\,573. The panel layout and model grid parameters are the same as in Figure~\ref{fig:5063_3d}, except that the fast-shock model adopts $\sigma_{[\rm O III]}=\frac23v_s$, following \citet{allen_mappings_2008} to account for thermally unstable shocks, whose behavior resembles that of steady-flow shocks with velocities of $\sim\frac23 v_{s}$.
\label{fig:573_3d}}
\end{figure*}

\begin{figure*}[hbt]
\begin{interactive}{js}{NGC424ppxf_int.zip}
\epsscale{1.13}
\plotone{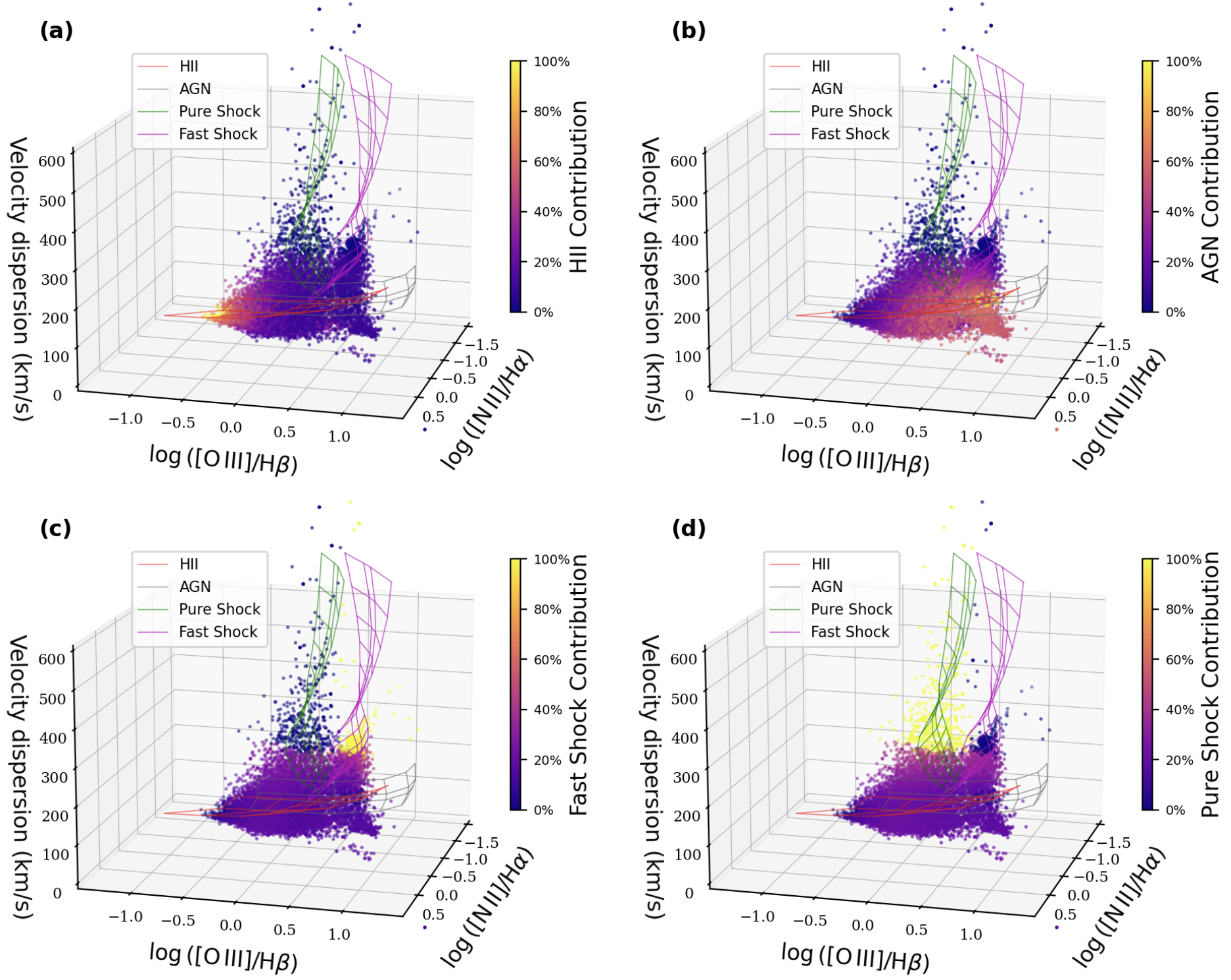}
\end{interactive}
\caption{Theoretical 3D diagrams of the MUSE IFU data for NGC\,424. The panel layout and model grid parameters are the same as in Figure~\ref{fig:5063_3d}\label{fig:424_3d}}
\end{figure*}

\begin{figure*}[hbt]
\begin{interactive}{js}{NGC1386ppxf_int.zip}
\epsscale{1.13}
\plotone{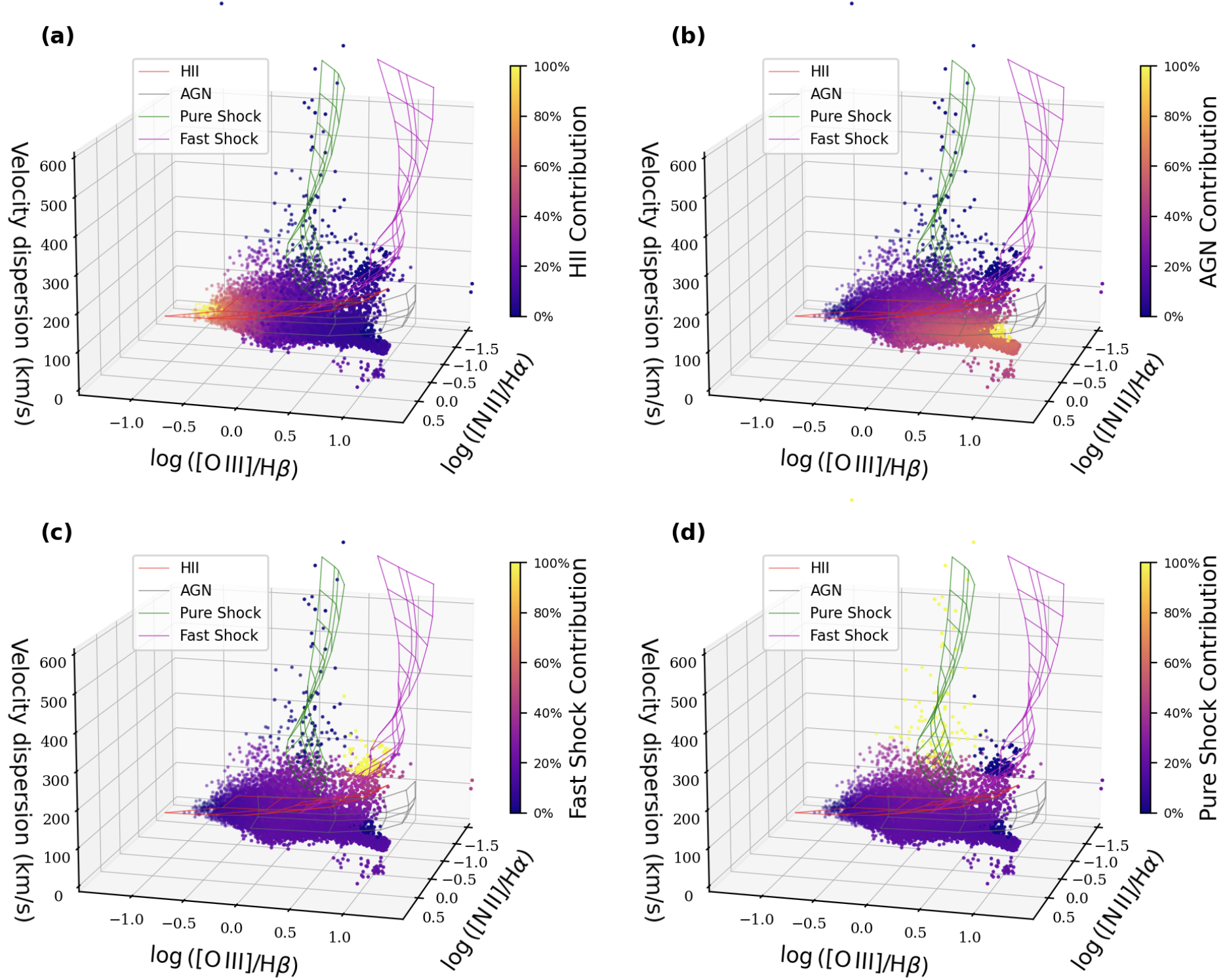}
\end{interactive}
\caption{Theoretical 3D diagrams of the MUSE IFU data for NGC\,1386. The panel layout and model grid parameters are the same as in Figure~\ref{fig:5063_3d}, except that the fast-shock model adopts $\eta_M=0.001$. 
\label{fig:1386_3d}}
\end{figure*}

\begin{figure*}[hbt]
\begin{interactive}{js}{NGC3081ppxf_int.zip}
\epsscale{1.13}
\plotone{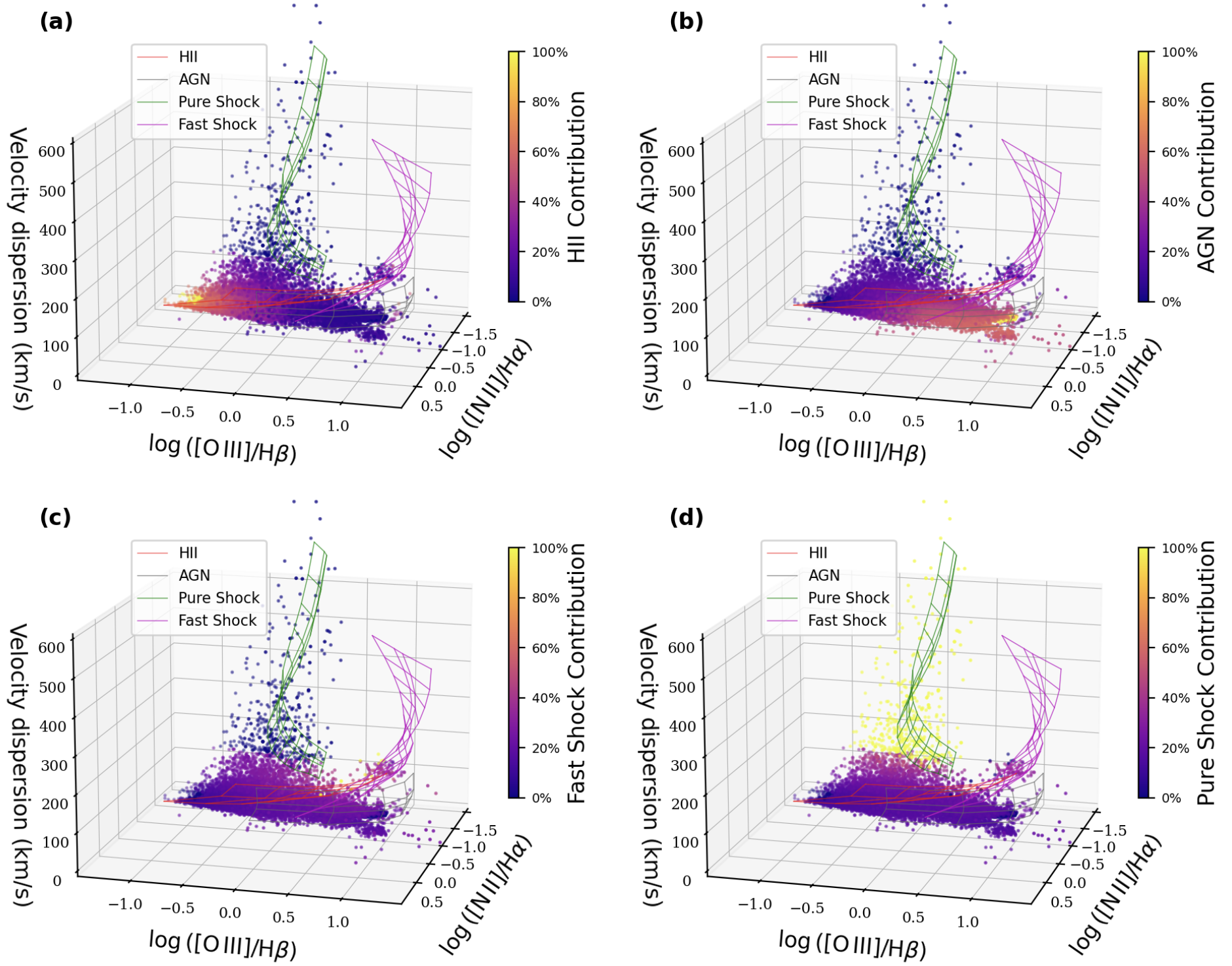}
\end{interactive}
\caption{Theoretical 3D diagrams of the MUSE IFU data for NGC\,3081. The panel layout and model grid parameters are the same as in Figure~\ref{fig:5063_3d}, except that the fast-shock model adopts $\sigma_{[\rm O III]}=\frac23v_s$, following \citet{allen_mappings_2008} to account for thermally unstable shocks, whose behavior resembles that of steady-flow shocks with velocities of $\sim\frac23 v_{s}$.
\label{fig:3081_3d}}
\end{figure*}

\begin{figure*}[htb]
\begin{interactive}{js}{NGC3393ppxf_int.zip}
\epsscale{1.13}
\plotone{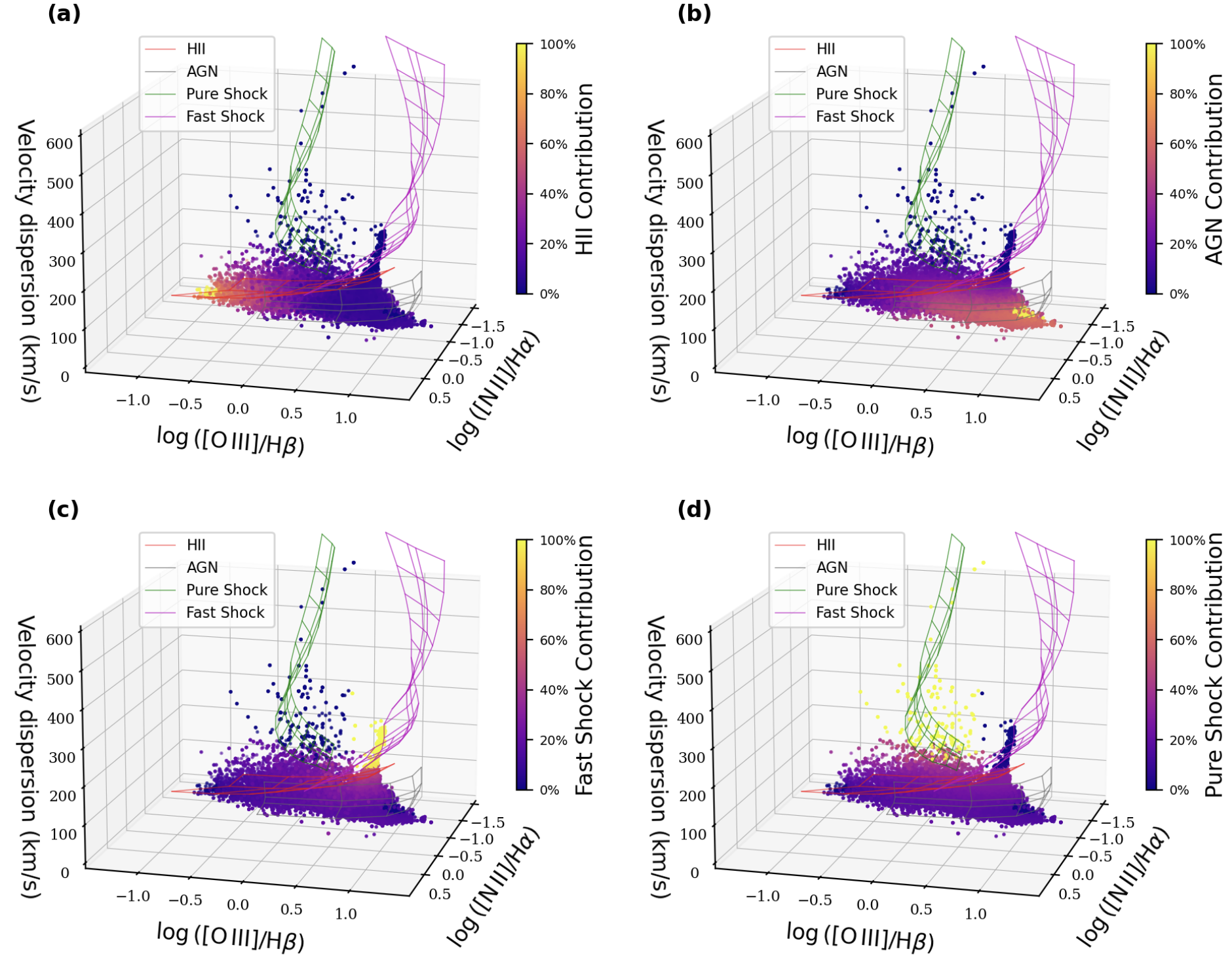}
\end{interactive}
\caption{Theoretical 3D diagrams of the MUSE IFU data for NGC\,3393. The panel layout and model grid parameters are the same as in Figure~\ref{fig:5063_3d}, except that the fast-shock model adopts $\eta_M=0.001$.
\label{fig:3393_3d}}
\end{figure*}

\begin{figure*}[hbt]
\begin{interactive}{js}{NGC5643ppxf_int.zip}
\epsscale{1.13}
\plotone{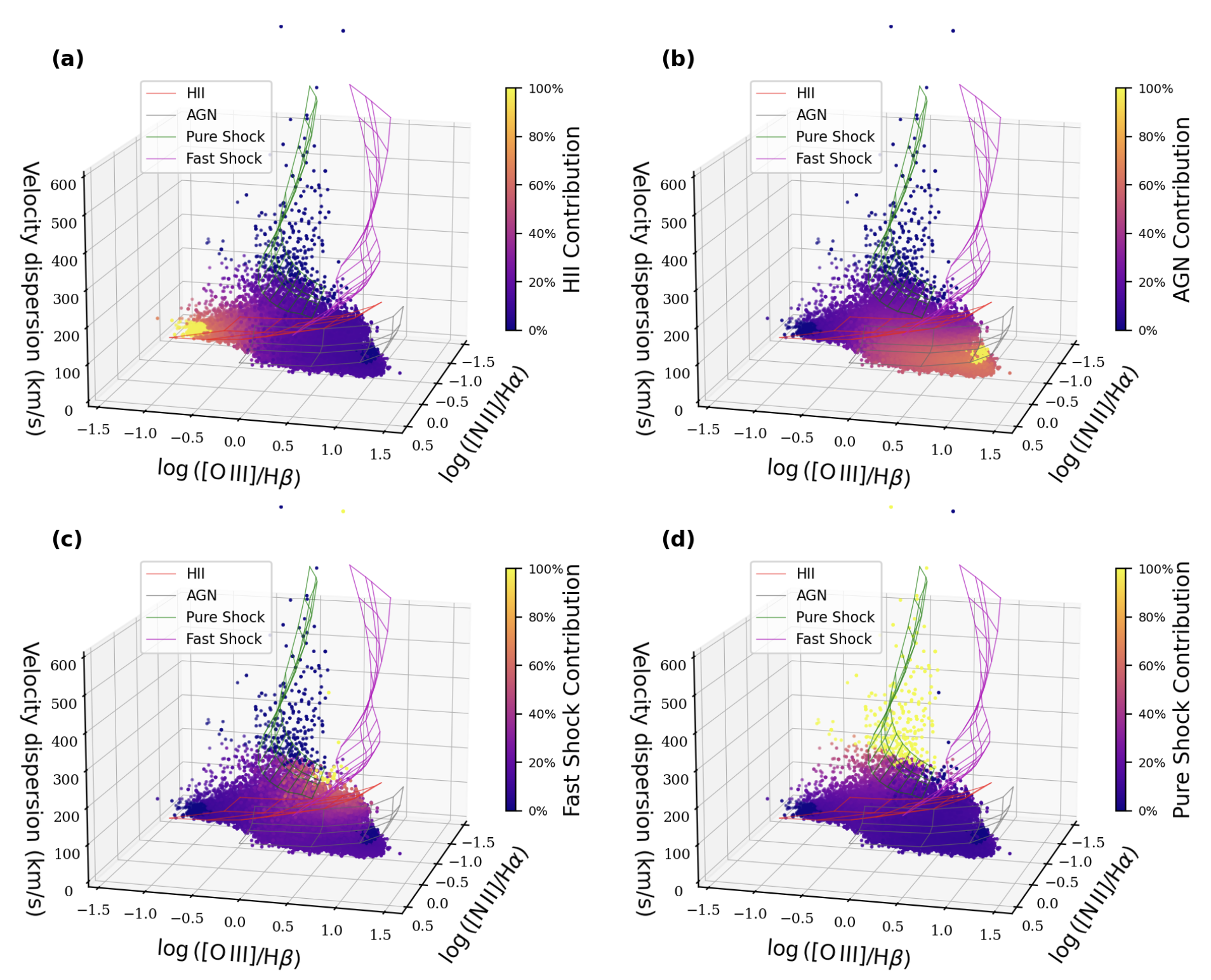}
\end{interactive}
\caption{Theoretical 3D diagrams of the MUSE IFU data for NGC\,5643. The panel layout and model grid parameters are the same as in Figure~\ref{fig:5063_3d}.
\label{fig:5643_3d}}
\end{figure*}

\begin{figure*}[hbt]
\begin{interactive}{js}{NGC7212ppxf_int.zip}
\epsscale{1.13}
\plotone{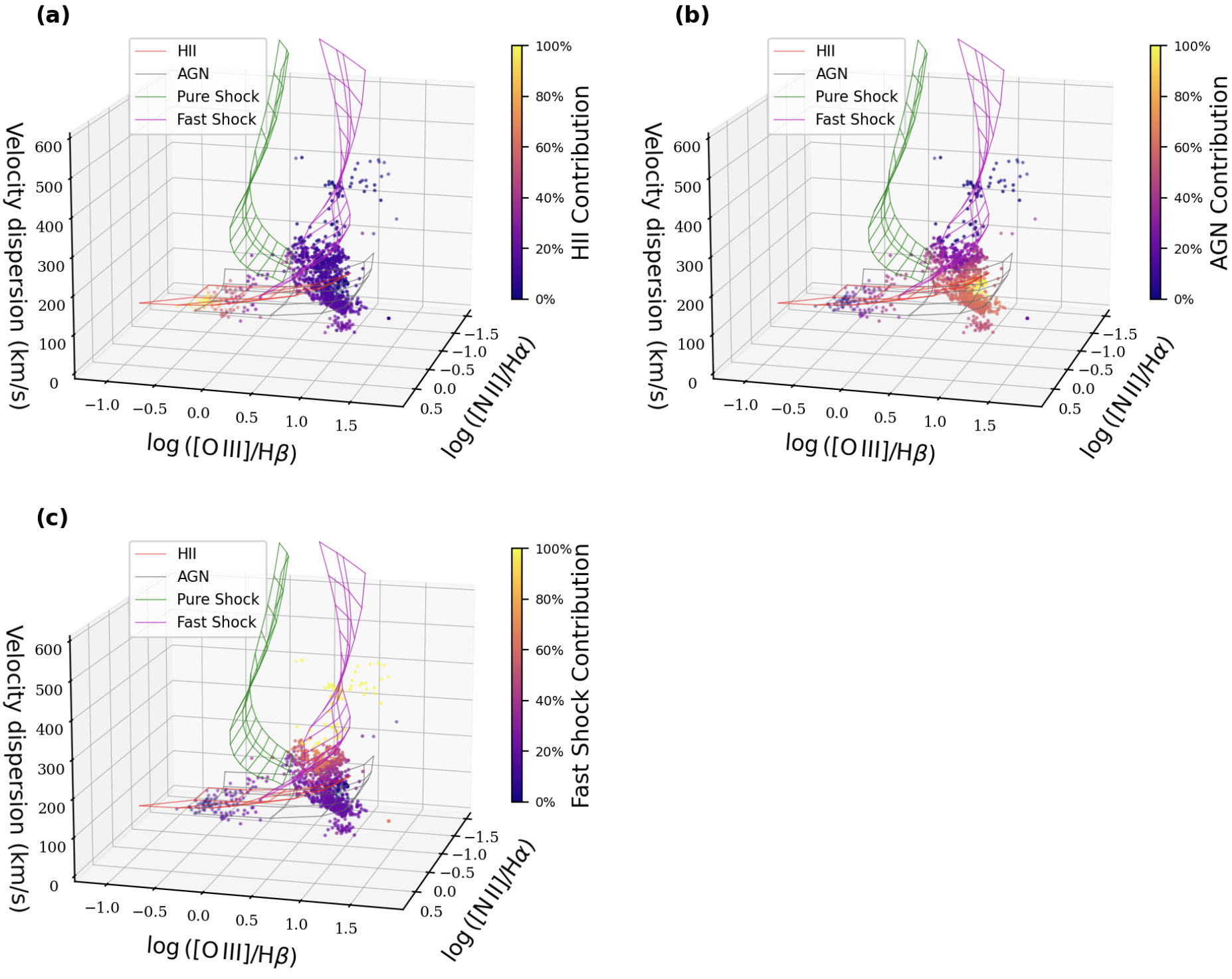}
\end{interactive}
\caption{Theoretical 3D diagrams of the MUSE IFU data for NGC\,7212. The panel layout and model grid parameters are the same as in Figure~\ref{fig:5063_3d}.
\label{fig:7212_3d}}
\end{figure*}

\setlength{\tabcolsep}{2pt}
\begin{deluxetable}{c|c|c|c|c}[hbt]
\centering
\tablewidth{2pc}
\tablecaption{Varying Parameters in Theoretical Models\label{tab:3}}
\tablenum{A1}
\tablehead{
  \colhead{Galaxy} &
  \colhead{$\sigma_{[\rm O III], HII}$} &
  \colhead{$\sigma_{[\rm O III], AGN}$} &
  \colhead{fast shock} &
  \colhead{pure shock}
  \\
  \colhead{} &
  \colhead{($\rm km\,s^{-1}$)} &
  \colhead{($\rm km\,s^{-1}$)} &
  \colhead{$L_{\mathrm{H}\beta,\mathrm{sh}} : L_{\mathrm{H}\beta,\mathrm{pre}}$} &
  \colhead{$L_{\mathrm{H}\beta,\mathrm{sh}} : L_{\mathrm{H}\beta,\mathrm{pre}}$}
  }
\startdata
IC 5063 & 81 & 123 & 0.65:0.35 & 0.95:0.05  \\
MRK 573 & 87 & 104 & 0.2:0.8 & 1.0:0.0 \\
NGC 424 & 77 & 169 & 0.7:0.3 & 0.95:0.05 \\
NGC1386 & 87 & 131 & 0.3:0.7 & 0.95:0.05 \\
NGC3081 & 85 & 101 & 0.2:0.8 & 1.0:0.0 \\
NGC3393 & 82 & 106 & 0.2:0.8 & 1.0:0.0 \\
NGC5643 & 85 & 116 & 0.5:0.5 & 1.0:0.0 \\
NGC7212 & 79 & 118 & 0.5:0.5 & 1.0:0.0 \\
\enddata
\end{deluxetable}

\clearpage
\section{The full-field separation maps for NGC3081 and NGC7212} \label{sec:appendixB}
\renewcommand{\thefigure}{B\arabic{figure}}
\setcounter{figure}{0}

\begin{figure*}[htb]
\epsscale{1.2}
\plotone{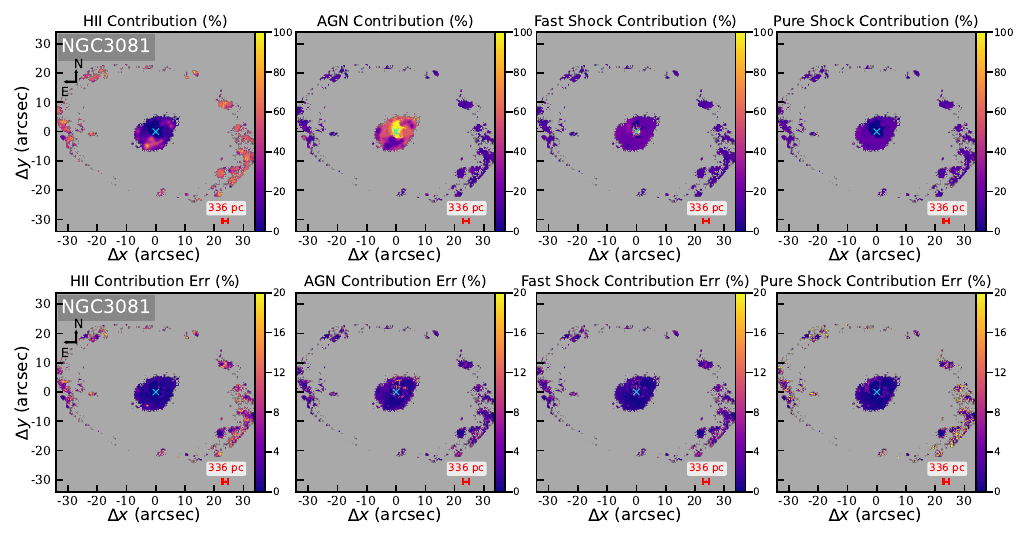}
\caption{The Full-field 2D fractional contribution maps for NGC\,3081.
\label{fig:A1}}
\end{figure*}

\begin{figure*}[h]
\epsscale{0.94}
\plotone{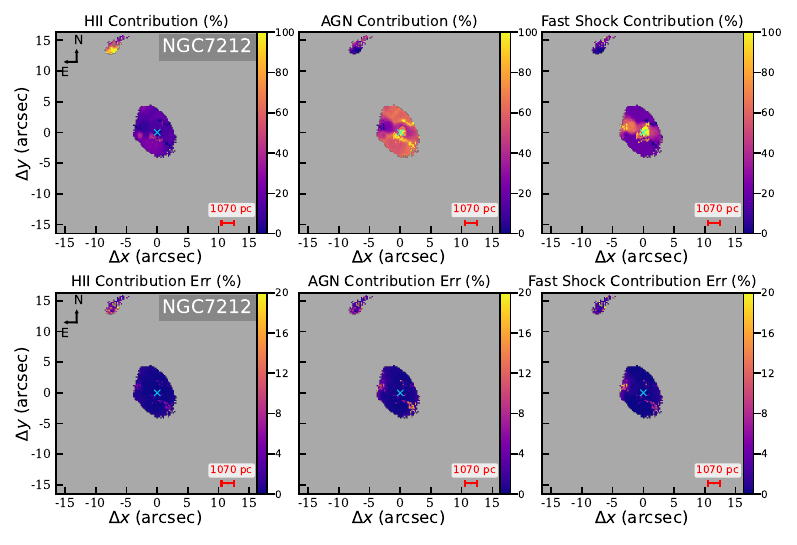}
\caption{The Full-field 2D fractional contribution maps for NGC\,7212.
\label{fig:A2}}
\end{figure*}

\end{document}